\documentclass[twocolumn,showpacs,preprintnumbers,amsmath,amssymb,showcolor]{revtex4}
\pdfoutput=1

\usepackage{graphicx}
\usepackage{dcolumn}
\usepackage{bm}
\usepackage{amssymb}
\usepackage{color}

 \def\tskip{\setlength{\tskip}{5pt}}
\def\colwidth{\setlength{\colwidth}{3.5in}}

\def\prd{Phys. Rev. D}

\def\apj{Astrophys. J.~}

\def\mnras{Mon. Not. Roy. Astron. Soc.~}
\def\apjs{Astrophys. J. Suppl. Ser.~}

\def\aap{A \& A}

\newcommand{\lsim}{\mathrel{\hbox{\rlap{\lower.55ex\hbox{$\sim$}} \kern-.3em \raise.4ex \hbox{$<$}}}}
\newcommand{\gsim}{\mathrel{\hbox{\rlap{\lower.55ex\hbox{$\sim$}} \kern-.3em \raise.4ex \hbox{$>$}}}}
\newcommand{\beq}{\begin{equation}}
\newcommand{\eeq}{\end{equation}}
\newcommand{\be}{\begin{equation}}
\newcommand{\ee}{\end{equation}}
\newcommand{\bes}{\begin{equation*}}
\newcommand{\ees}{\end{equation*}}
\newcommand{\beqa}{\begin{eqnarray}}
\newcommand{\eeqa}{\end{eqnarray}}
\newcommand{\bea}{\begin{eqnarray}}
\newcommand{\ena}{\end{eqnarray}}

\begin{document}

\title{Probing CMB Cold Spot through Local Minkowski Functionals}

\author{Wen Zhao}
\affiliation{Key Laboratory for Researches in Galaxies and Cosmology, Department of Astronomy,
University of Science and Technology of China, Hefei, 230026,
China}

%\date{\today}

%%%%%%%%%%%%%%%%%%%%%%%%%%%%%%%%%%%%%%%%%%%%%%%%%%%%%%%%%%%%%%%%%%%%%%%%%%%%%%%%%%%
%%%%%%%%%%%%%%%%%%%%%%%%%%%%%%%%%%  ABSTRACT  %%%%%%%%%%%%%%%%%%%%%%%%%%%%%%%%%%%%%%%%%%
%%%%%%%%%%%%%%%%%%%%%%%%%%%%%%%%%%%%%%%%%%%%%%%%%%%%%%%%%%%%%%%%%%%%%%%%%%%%%%%%%%%
\begin{abstract}

Both WMAP and PLANCK missions reported the extremely Cold Spot
(CS) centered at Galactic coordinate ($l=209^{\circ}$,
$b=-57^{\circ}$) in CMB map. In this paper, we study the local
non-Gaussianity of CS by defining the local Minkowski functions.
We find that the third Minkowski function $\nu_2$ is quite
sensitive to the non-Gaussianity caused by CS. Compared with the
random Gaussian simulations, WMAP CS deviates from Gaussianity at
more than $99\%$ confident level at the scale $R\sim10^{\circ}$.
Meanwhile, we find that cosmic texture provides an excellent
explanation for these anomalies related to WMAP CS, which could be
further tested by the future polarization data.

\end{abstract}

%%%%%%%%%%%%%%%%%%%%%%%%%%%%%%%%%%%%%%%%%%%%%%%%%%%%%%%%%%%%%%%%%%%%%%%%%%%%%%%%%%%
%%%%%%%%%%%%%%%%%%%%%%%%%%%%%%%%%%%%%%%%%%%%%%%%%%%%%%%%%%%%%%%%%%%%%%%%%%%%%%%%%%%


\maketitle

%%%%%%%%%%%%%%%%%%%%%%%%%%%%%%%%%%%%%%%%%%%%%%%%%%%%%%%%%%%%%%%%%%%%%%%%%%%%%%%%%%%
%%%%%%%%%%%%%%%%%%%%%%%%%%%%%%%%%%  SECTION 1   %%%%%%%%%%%%%%%%%%%%%%%%%%%%%%%%%%%%%%%%%%
%%%%%%%%%%%%%%%%%%%%%%%%%%%%%%%%%%%%%%%%%%%%%%%%%%%%%%%%%%%%%%%%%%%%%%%%%%%%%%%%%%%

\section{Introduction}

Soon after the release of observations of the NASA Wilkinson
Microwave Anisotropy Probe (WMAP) satellite on the Cosmic
Microwave Background (CMB) temperature and polarization
anisotropies, some anomalies in CMB field have also been reported
\cite{bennett2011}. Among these, an extremely Cold Spot (CS)
centered at Galactic coordinate ($l=209^{\circ}$, $b=-57^{\circ}$)
and a characteristic scale about $10^{\circ}$ was detected in the
Spherical Mexican Hat Wavelet (SMHW) non-Gaussian analysis
\citep{vielva2004}. Comparing with the distribution derived from
the isotropic and Gaussian CMB simulations, due to this CS, the
SMHW coefficients of WMAP data have an excess of kurtosis
\cite{cruz2005}. The non-Gaussian CS has also been confirmed by
using some other statistics
\cite{cruz2007a,cruz2005,cayon2005,naselsky2010,zhang2010,vielva2010}.
At the same time, some analyses on WMAP CS have also been done,
such as the non-Gaussian tests for the different detectors and
different frequency channels of WMAP satellite
\cite{vielva2004,cruz2005}, the investigation of the NVSS sources
\cite{rudnick2007,smith2010}, the survey around the CS with
MegaCam on the Canada-France-Hawaii Telescope \cite{granett2009},
the redshift survey using VIMOS on VLT towards CS
\cite{bremer2010}, and the cross-correlation between WMAP and
Faraday depth rotation map \cite{hansen2012}.

Since then, various alternative explanations for this CS have been
proposed, including the possible foregrounds
\cite{cruz2006,hansen2012}, Sunyaev-Zeldovich effect
\cite{cruz2008}, the supervoid in the Universe
\cite{inoue2006,inoue2007,inoue2012}, and the cosmic texture
\cite{cruz2007b,cruz2008}. Due to the fact that, nearly all the
explanations of CS are related to the local characters of the CMB
field, the studies on the local properties of CS are necessary. In
the previous work \cite{zhao2012}, we have studied the local
non-Gaussian properties of WMAP CS by the local mean temperature,
variance, skewness and kurtosis. We found the excesses of the
local variance and skewness in the large scales with
$R>5^{\circ}$, which implies that WMAP CS is a large-scale
non-Gaussian structure, rather than a combination of some small
structures. In this paper, we shall focus on the same topic by
using the different local statistics. We introduce the local
Minkowski Functions (MFs), and apply them to the WMAP data, in
particular WMAP CS. Comparing with random Gaussian simulation, we
find that for the statistics based on MF $\nu_2$, WMAP CS
significantly deviates from Gaussianity at large scale, especially
at the scale $R\sim 10^{\circ}$. Meanwhile, similar to
\cite{zhao2012}, we find that these local non-Gaussianities of
WMAP CS can be excellently explained by a cosmic texture. If
subtracting this texture from WMAP data, we find that all these
anomilies of MFs disappear. So our local analysis of the CS in
this paper also strongly supports the cosmic texture explanation.

The outline of this paper is as follows.
In Sec. \ref{sec2}, we introduce the WMAP data, which will be used in the analysis.
In Sec. \ref{sec3}, we define the local MFs and apply them into WMAP data.
In Section \ref{sec4}, we summarize the main results of this paper.

\section{The WMAP data}
\label{sec2} In our analysis, we shall use the WMAP data including
the ILC7 map and the NILC5 map. The WMAP instrument is composed of
10 DAs spanning five frequencies from 23 to 94 GHz
\cite{bennett2003}. The internal linear combination (ILC) method
has been used by WMAP team to generate the ILC maps
\cite{hinshaw2007,gold2011}. The 7-year ILC (written as ``ILC7")
map is a weighted combination from all five original frequency
bands, which are smoothed to a common resolution of one degree.
For the 5-year WMAP data, in \cite{delabrouille2009} the authors
have made a higher resolution CMB ILC map (written as ``NILC5"),
an implementation of a constrained linear combination of the
channels with minimum error variance on a frame of spherical
called needlets \cite{note1}. In this paper, we will consider both
these two maps for the analysis. Note that these WMAP data have
the same resolution parameter $N_{\rm side}=512$, and the
corresponding total pixel number $N_{\rm pix}=3145728$.

In comparison with WMAP observations to give constraint on the
statistics, a ${\rm \Lambda CDM}$ cosmology is assumed with the
cosmological parameters given by the WMAP 7-year best-fit values
\cite{komatsu2011}: $100\Omega_{\rm b}h^2=2.255$,
$\Omega_ch^2=0.1126$, $\Omega_{\rm \Lambda}=0.725$, $n_s=0.968$,
$\tau=0.088$ and $\Delta^2_{\mathcal{R}}(k_0)=2.430\times10^{-9}$
at $k_0=0.002{\rm Mpc}^{-1}$. Similar to the previous work
\citep{hansen2012,zhao2012}, to simulate the ILC7 map, we ignore
the noises and smooth the simulated map with one degree
resolution. And for NILC5, we consider the noise level and beam
window function given in \cite{delabrouille2009}. In all the
random Gaussian simulations, we assume that the temperature
fluctuations and instrument noise follow the Gaussian
distribution, and do not consider any effect due to the residual
foreground contaminations.

\section{Applying Local Minkowski Functions to WMAP data}
\label{sec3}
\subsection{Minkowski Functions}
MFs characterize the morphological properties of convex, compact
sets in an $n$-dimensional space. On the 2-dimensional spherical
surface $\mathbb{S}^2$, any morphological property can be expanded
as a linear combination of three MFs, which represent the area,
circumference and integrated geodesic curvature of an excursion
set \cite{schmalzing1998}. For a given threshold $\nu$, it is
convenient to define the excursion set $Q_{\nu}$ and its boundary
$\partial Q_{\nu}$ of a smooth scalar field $u$ as follows:
$Q_{\nu}=\left\{x\in \mathbb{S}^2|u(x)>\nu\right\}$ and $\partial
Q_{\nu}=\left\{x\in \mathbb{S}^2|u(x)=\nu\right\}$. Then, the MFs
$v_0$, $v_1$ and $v_2$ can be written as \cite{schmalzing1998},
 \begin{equation}
 v_0(\nu):=\int_{Q_{\nu}} \frac{{\rm d}a}{4\pi}, ~v_1:=\int_{\partial Q_{\nu}} \frac{{\rm d}\ell}{16\pi}, ~v_2:=\int_{\partial Q_{\nu}} \frac{{\rm d}\ell~\kappa}{8\pi^2},
 \end{equation}
where ${\rm d}a$ and ${\rm d}\ell$ denote the surface element of
${\rm \mathbb{S}^2}$ and the line element along
$\partial{Q_{\nu}}$, respectively. And $\kappa$ is the geodesic
curvature. Given a pixelated map with field $u(x_i)$, these MFs
can be numerically calculated by the formulae
\citep{schmalzing1998,lim2012}
 \begin{equation}
 v_i(\nu)=\frac{1}{N_{\rm pix}} \sum_{k=1}^{N_{\rm pix}} \mathcal{I}_i(\nu,x_k),~~(i=0,1,2),\label{eq2}
 \end{equation}
where
% \begin{eqnarray}
% \mathcal{I}_0(\nu,x_k)&:=&\Theta(u-\nu), \nonumber \\
% \mathcal{I}_1(\nu,x_k)&:=&\frac{\delta(u-\nu)}{4}\sqrt{u_{;\theta}^2+u_{;\phi}^2}, \label{eq3}\\
% \mathcal{I}_2(\nu,x_k)&:=&\frac{\delta(u-\nu)}{2\pi}\frac{2u_{;\theta}u_{;\phi}u_{;\theta\phi}-u_{;\theta}^2u_{;\phi\phi}-u_{;\phi}^2u_{;\theta\theta}}{u_{;\theta}^2+u_{;\phi}^2}. \nonumber
% \end{eqnarray}
 \begin{eqnarray}
 \mathcal{I}_0(\nu,x_k)&:=&\Theta(u-\nu), \nonumber \\
 \mathcal{I}_1(\nu,x_k)&:=&\frac{\delta(u-\nu)}{4}\mathcal{U}_1(x_k), \nonumber\\
 \mathcal{I}_2(\nu,x_k)&:=&\frac{\delta(u-\nu)}{2\pi}\mathcal{U}_2(x_k), \label{eq3} \\
 \mathcal{U}_1(x_k)    &:=&\sqrt{u_{;\theta}^2+u_{;\phi}^2}, \nonumber \\
 \mathcal{U}_2(x_k)    &:=&\frac{2u_{;\theta}u_{;\phi}u_{;\theta\phi}-u_{;\theta}^2u_{;\phi\phi}-u_{;\phi}^2u_{;\theta\theta}}{u_{;\theta}^2+u_{;\phi}^2}. \nonumber
 \end{eqnarray}
Note that $u_{;i}$ denotes the covariant differentiation of $u$
with respect to the coordinate $i$. The delta function in these
formulae can be numerically approximated through a discretization
of threshold space in bins of width $\Delta\nu$ by the
stepfunction
$\delta_N(x)=(\Delta\nu)^{-1}[\Theta(x+\Delta\nu/2)-\Theta(x-\Delta\nu/2)]$.
The expectation values of three MFs for a Gaussian random field
are also derived in \cite{tomita1986}, which have been explicitly
expressed in equations (14) and (15) in \cite{schmalzing1998}.

These MFs have been applied by cosmologists to look for
derivations from Gaussianity of the perturbations in the CMB
\cite{winitzki1998,schmalzing1998,novikov1999,eriksen2004,hikage2006,hikage2008,hikage2009,hikage2012,komatsu2009,matsubara2010}.
In particular, in \cite{lim2012} the authors used the MFs to probe
the cold/hot disk-like structure in the CMB. However, they found
these statistics are noise-dominated for WMAP CS. For a single
WMAP or PLANCK resolution map, the method can only detect the
highly prominent disk, i.e. extremely cold or hot disk with quite
large area. These can be easily understood as follows: since the
MFs are constructed on the global sky, the local non-Gaussianity
of the cold/hot spot in the relatively small scale has been
diluted to be too small.

\subsection{Local Minkowski Functions}
To study the local properties of WMAP CS, similar to the previous
works \cite{local1,local2,local3,zhao2012}, in this paper we shall
define the local MFs as the statistics of CMB field. For a given
full-sky WMAP data with $N_{\rm side}=512$ (ILC7 or NILC5), we
smooth them using a Gaussian filter with a smoothing scale of
$\theta_s$. Since MFs are sensitive to the smoothing scale of a
density field and thereby we can obtain a variety of information
from density fields by using different smoothing levels. In this
paper, we focus on both ILC fields smoothed by six different
smoothing scales $10'$, $20'$, $30'$, $40'$, $50'$, $60'$. Then,
we can construct the corresponding full-sky maps:
$\mathcal{U}_1(x_k)$ and $\mathcal{U}_2(x_k)$ defined in Eq.
(\ref{eq3}), where $u$ is the corresponding ILC temperature
anisotropic map.

Now, we can define the local MFs. Let $\Omega(\theta_j,\phi_j;R)$
be a spherical cap with aperture of $R$ degree, centered at
$(\theta_j,\phi_j)$. The local MFs $v_{i}(\nu)$ ($i=0,1,2$) in
this cap can be calculated by using Eq. (\ref{eq2}), which are
denoted as $v_i^j(\nu;R)$ in the rest of this paper. But here the
summation is carried out only for the pixels inside the cap
$\Omega(\theta_j,\phi_j;R)$. Note that in our calculation, the
binning range of threshold $\nu$ is set to be -3.0 to 3.0 with 24
equally spaced bins of $\nu/\sigma$ ($\sigma$ is the standard
deviation of $u$-field in this cap) per each MF. In order to
quantify the same kind of MFs by a single quantity, following
\cite{hikage2009}, for each $i$ (the type of MFs), \{$j$, $R$\}
(the cap) we can define the $\chi^2$ as follows:
 \begin{eqnarray}
 \chi^2=\sum_{\alpha\alpha'}[v_i^j(\nu_{\alpha};R)-v_i^{\rm th}(\nu_{\alpha};R)]\Sigma_{\alpha\alpha'}^{-1} \nonumber \\
  \times [v_i^{j}(\nu_{\alpha'};R)-v_i^{\rm th}(\nu_{\alpha'};R)], \label{chi2}
 \end{eqnarray}
where $\alpha$ and $\alpha'$ denote the binning number of
threshold values. $v_i^{\rm th}(\nu;R)$ is the theoretical value
of $v_i^j(\nu;R)$, which is independent of the superscript $j$.
$\Sigma$ is the corresponding covariance matrix. Although, in
principle the theoretical values $v_i^{\rm th}(\nu;R)$ for the
random Gaussian field can be calculated by the analytical formulae
\cite{tomita1986,schmalzing1998}, there are systematical
differences from the numerical results due to the binning of
threshold \cite{lim2012}. In this paper, we avoid this problem by
replacing the theoretical value $v_i^{\rm th}(\nu;R)$ by the
quantity $\langle{v_i(\nu;R)}\rangle$, which is the average value
of all ${v_i^j(\nu;R)}$ with $|b_j|>30^{\circ}$. And the
covariance matrix $\Sigma$ can also be numerically calculated by
these quantities ${v_i^j(\nu;R)}$. Note that the Galactic plane
have been excluded to reduce the effect of foreground residuals
\cite{note2}. Hereafter, we denote this $\chi^2$ quantity as
$X_i^j(R)$. Clearly, the values $X_i^j(R)$ obtained in this way
for each cap can be viewed as a measure of non-Gaussianity in the
direction of the center of cap $(\theta_j, \phi_j)$. For a given
aperture $R$, we scan the celestial sphere with evenly distributed
spherical caps, and build the $X_0(R)$-, $X_1(R)$-, $X_2(R)$-maps.
In our analysis, we have chosen the locations of centroids of
spots to be the pixels in $N_{\rm side}=64$ resolution. By
choosing different $R$ values, one can study the local properties
of CMB field at different scales.

Same to the $V(R)$-map (or $S(R)$-, $K(R)$-maps) defined in
\cite{zhao2012}, here we also find that $X_i^j(R)$ always maximize
at the edge of the circles, rather than the center of circles. To
overcome this problem and localize the non-Gaussian sources, we
define the average quantities
 \begin{equation}
 \bar{X}_i^j(R):=\frac{1}{N_{\rm pix}}\sum_{j=1}^{N_{\rm pix}} {X}_i^j(R),~~~(i=0,1,2),
 \end{equation}
where $N_{\rm pix}$ is again the pixel number in the $j^{\rm th}$ cap.

\begin{figure}
  \begin{center}
    \centerline{\includegraphics[scale=.10]{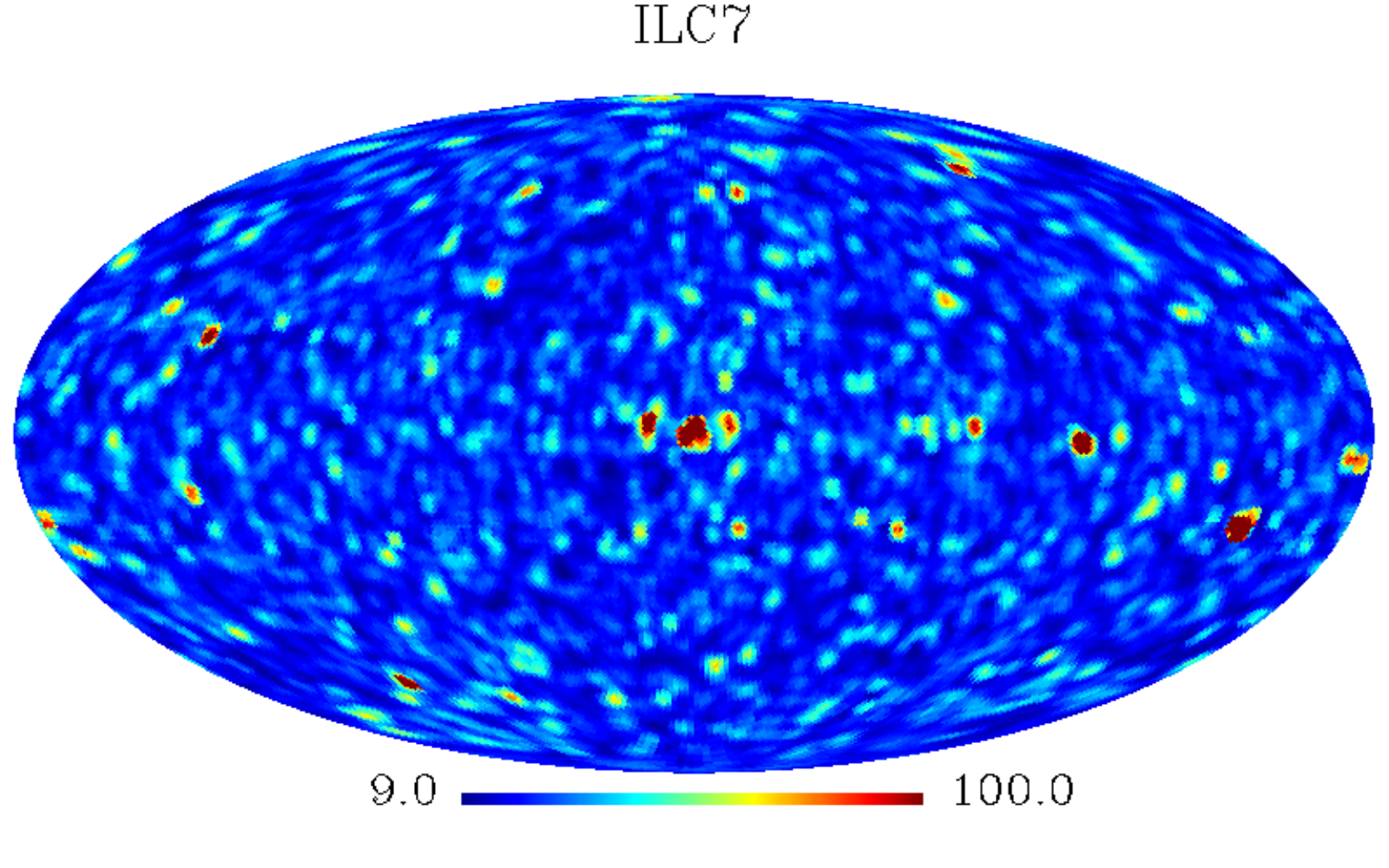}\includegraphics[scale=.10]{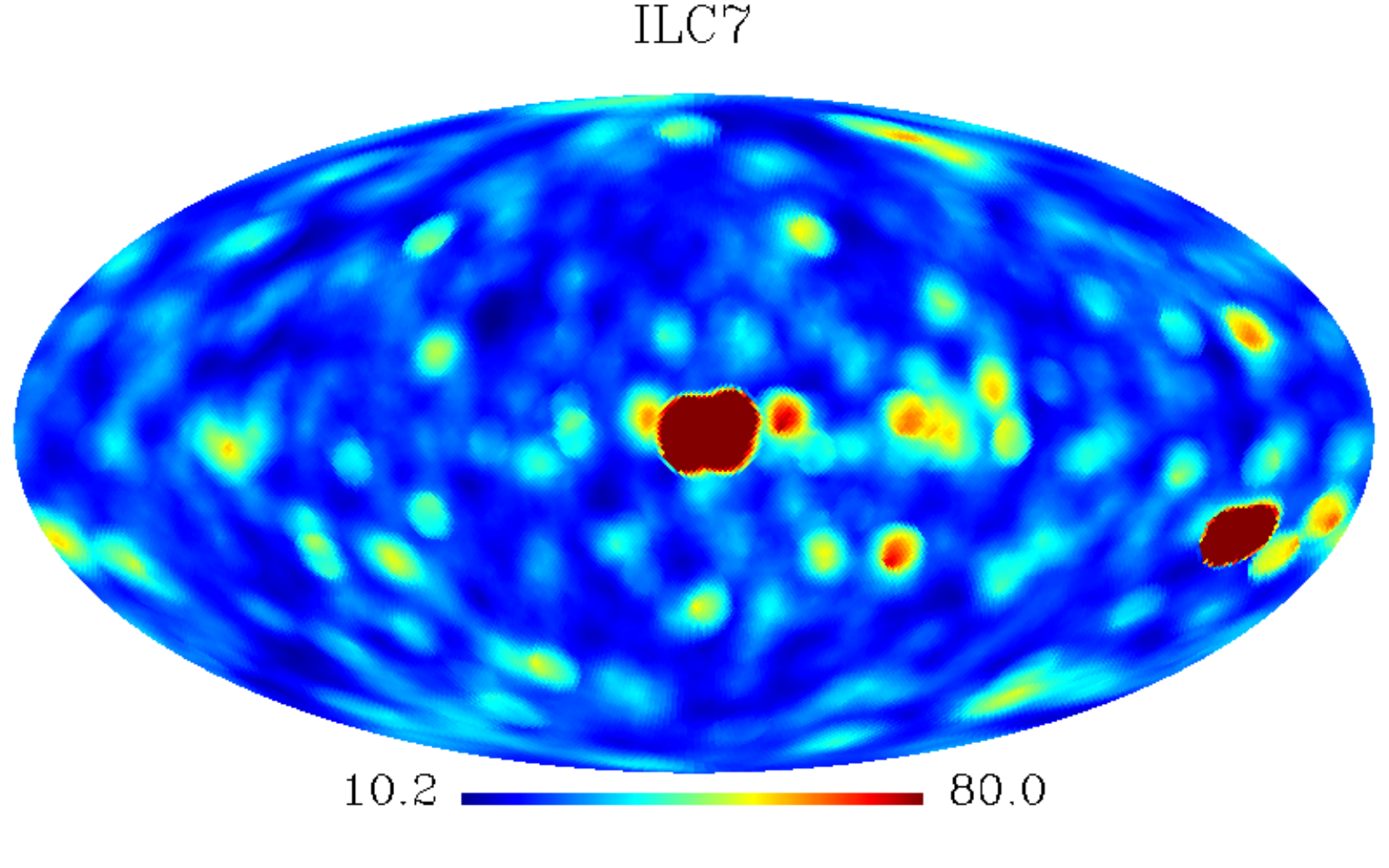}\includegraphics[scale=.10]{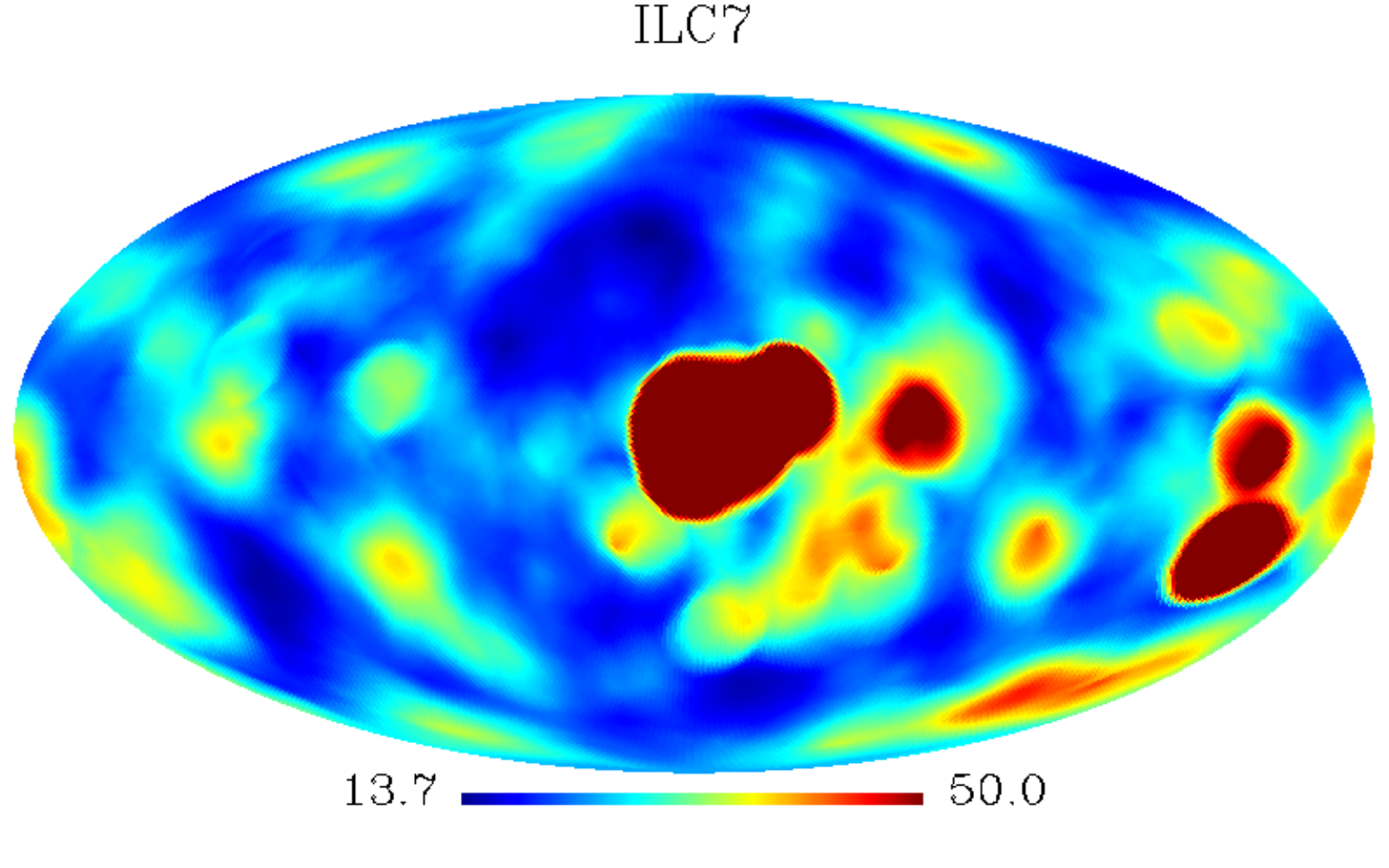}}
    \centerline{\includegraphics[scale=.10]{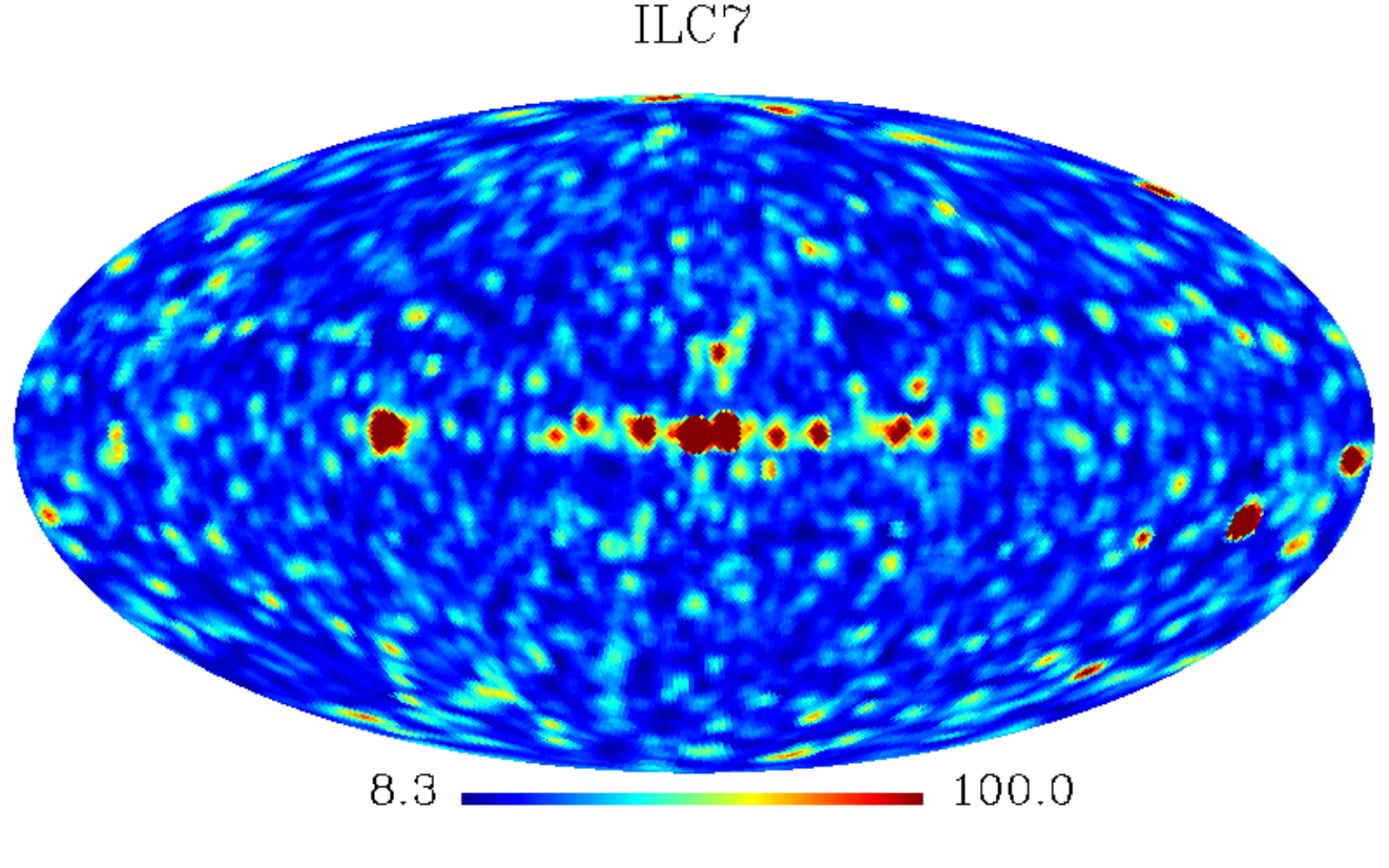}\includegraphics[scale=.10]{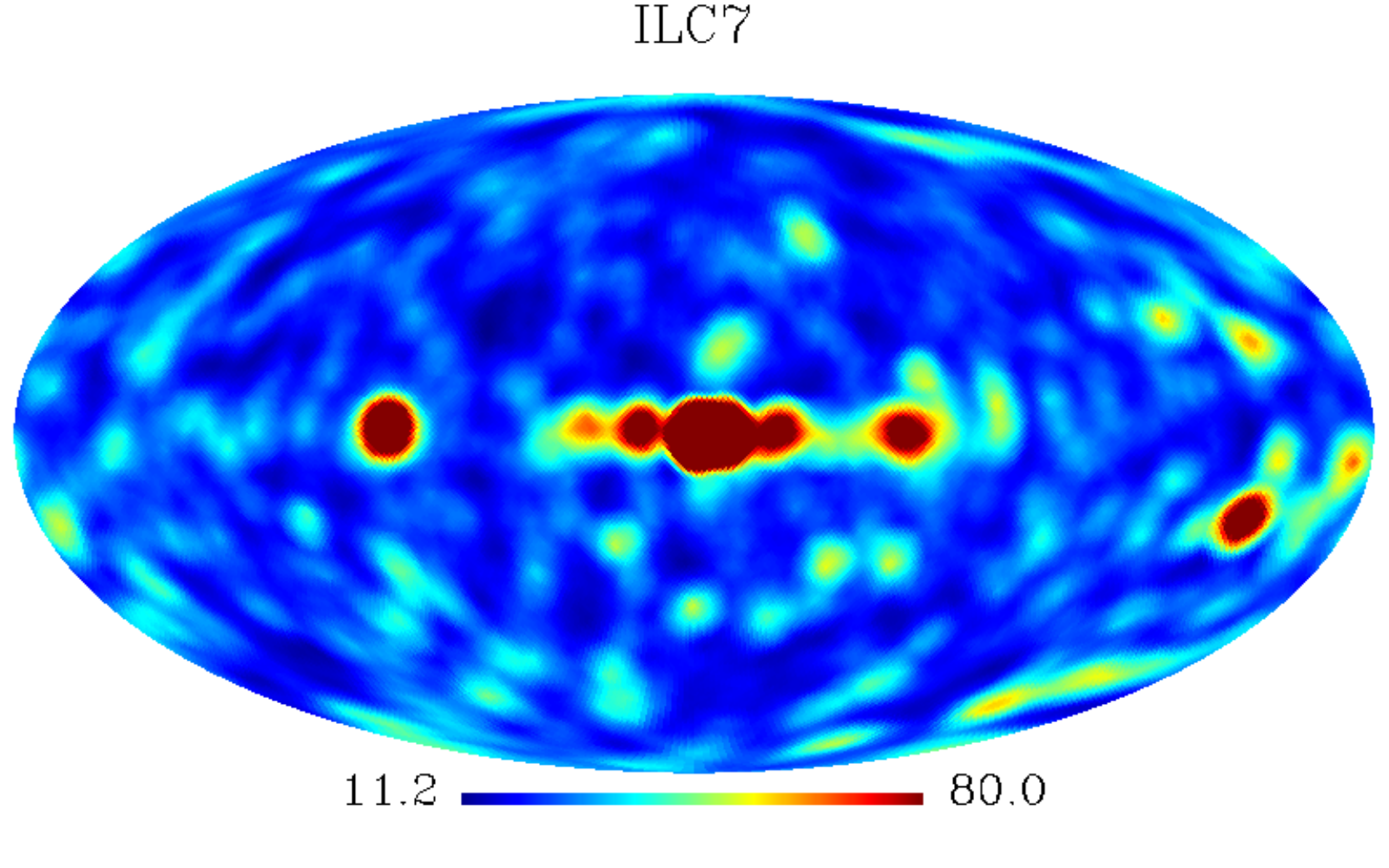}\includegraphics[scale=.10]{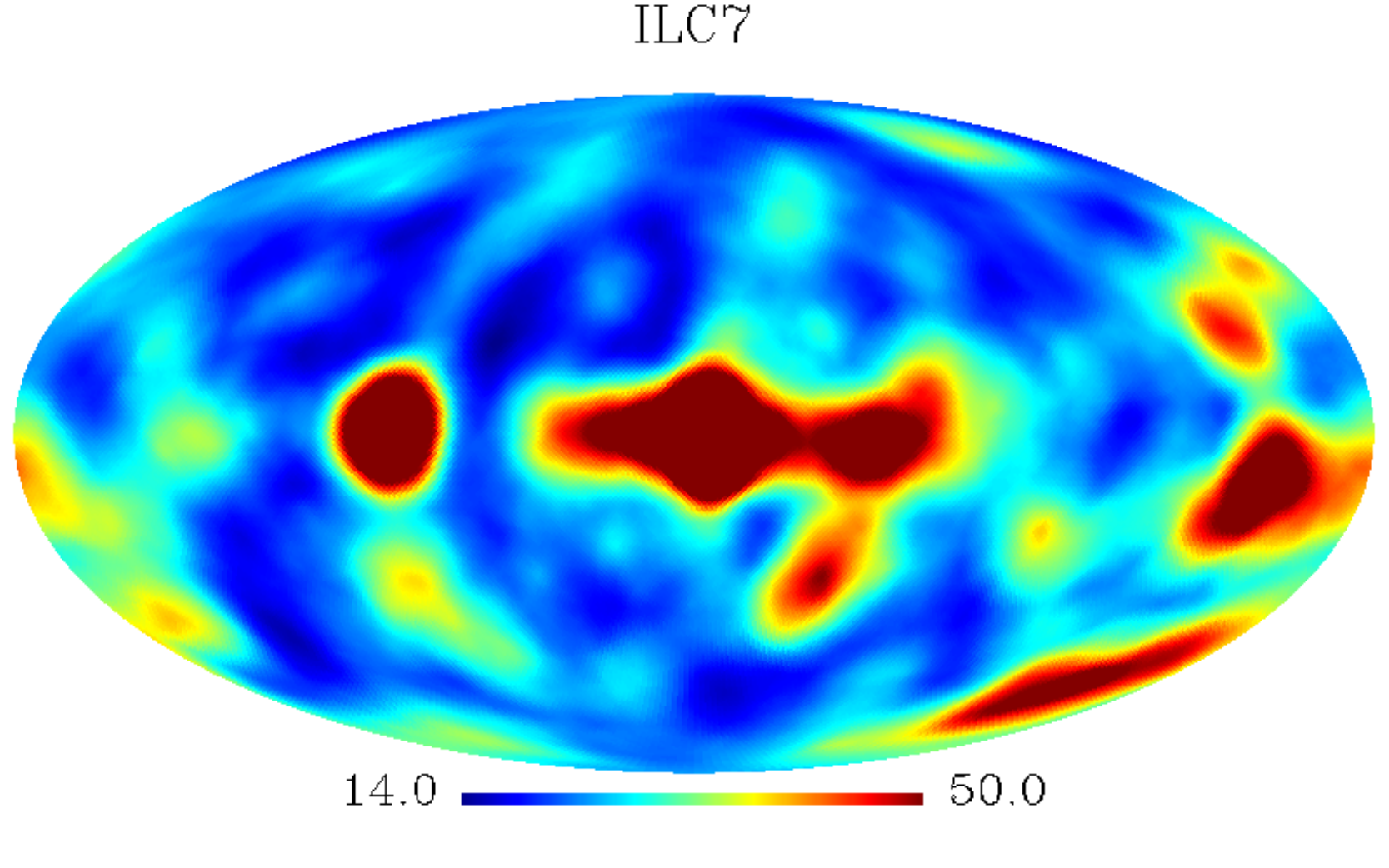}}
    \centerline{\includegraphics[scale=.10]{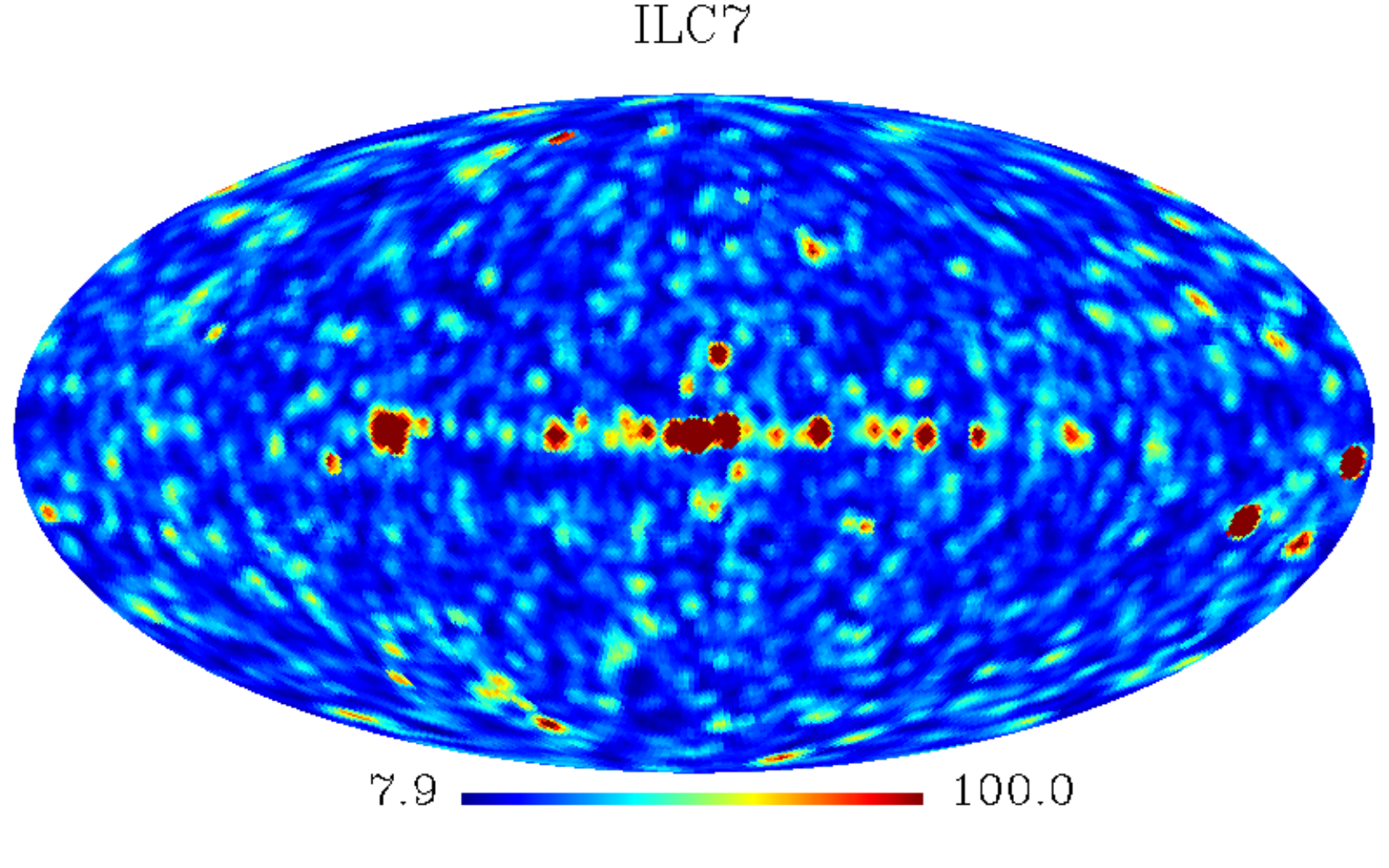}\includegraphics[scale=.10]{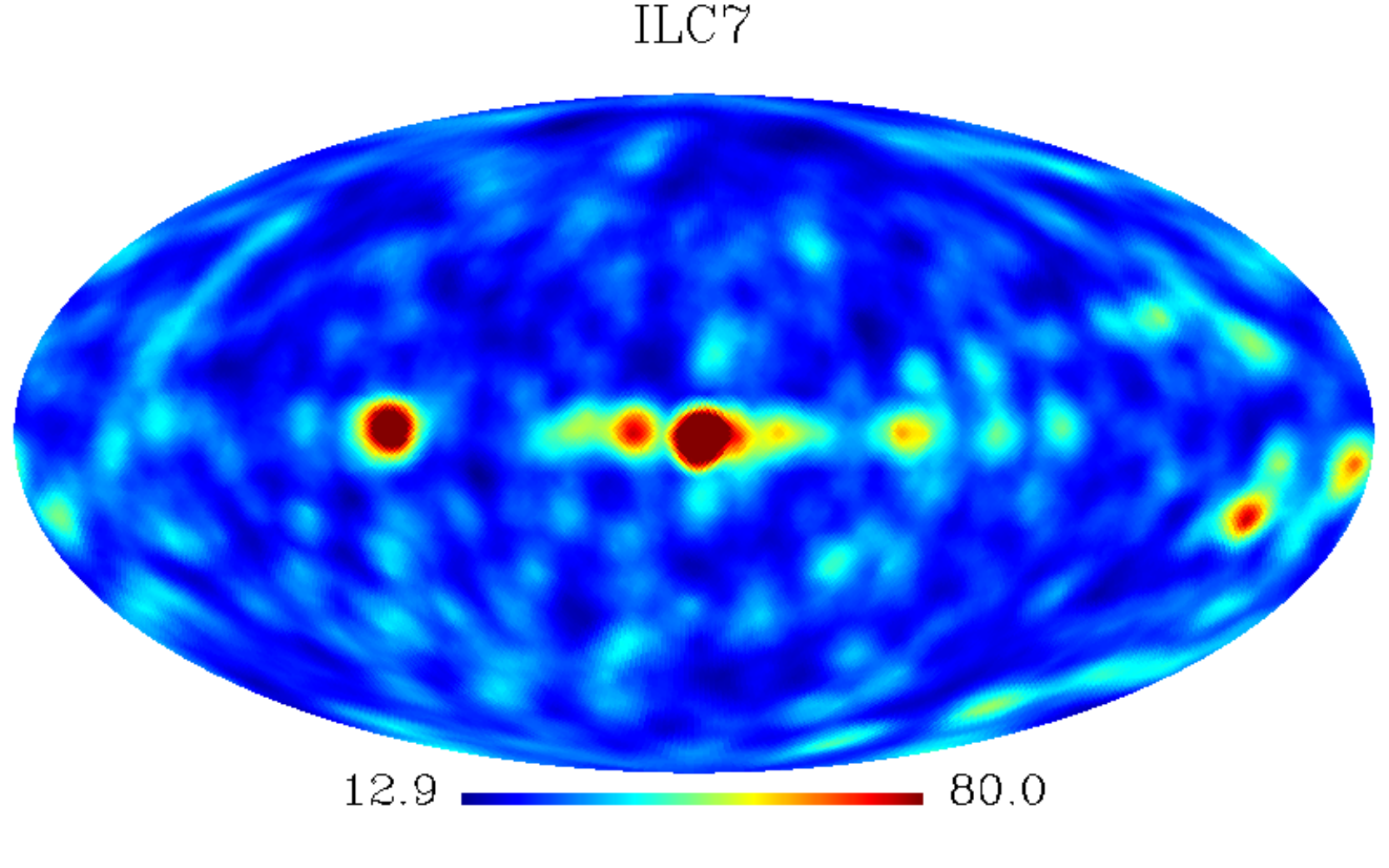}\includegraphics[scale=.10]{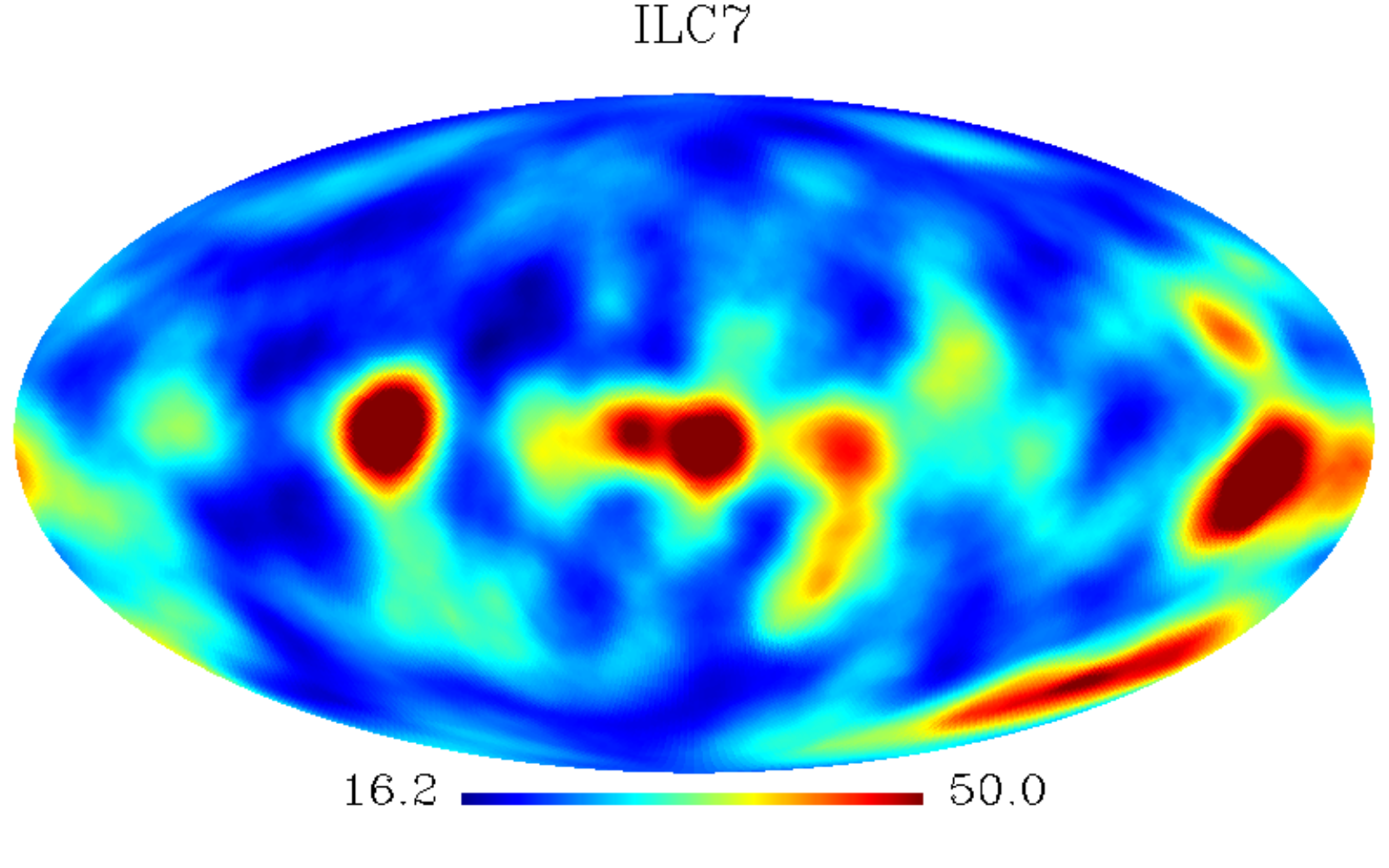}}
    \caption{$\bar{X}_0(R)$ maps (upper), $\bar{X}_1(R)$ maps (middle), and $\bar{X}_2(R)$ maps (lower) for ILC7 data with $\theta_s=60'$ smoothing.
    In the left panels, we have used $R=2^{\circ}$, in middle panels, $R=5^{\circ}$ is chosen, and in right panel, $R=10^{\circ}$.}
    \label{fig1}
  \end{center}
\end{figure}

We apply the method to the ILC7 data by choosing $R=2^{\circ}$ and
$\theta_s=60'$. The $\bar{X}_i$ maps are presented in Fig.
\ref{fig1} (left panels), which clearly show that these local
statistics, in particular third MF $v_2$, are very sensitive to
the foreground residuals and various point sources. The Galactic
plane is clearly presented, which is the non-Gaussian area caused
by the foreground residuals in ILC7 map. In addition, two
important point sources at ($l=209.5^{\circ}$, $b=-20.1^{\circ}$)
and ($l=184.9^{\circ}$, $b=-5.98^{\circ}$), as well as several
small ones, are also clearly found in the $\bar{X}_i$ maps. So we
expect these local statistics with small $R$ values can be used to
identify the point sources and foreground residuals, which will be
discussed in a separate paper. If we choose $R=5^{\circ}$, from
the middle panels in Fig. \ref{fig1} we find the similar results,
except for some non-Gaussianities of small point sources, which
have been diluted for this larger $R$ case. In the right panels,
we have chosen $R=10^{\circ}$, where the significant
non-Gaussianity around WMAP CS are clearly presented in all three
maps.

\begin{figure}
  \begin{center}
    \centerline{\includegraphics[scale=.15]{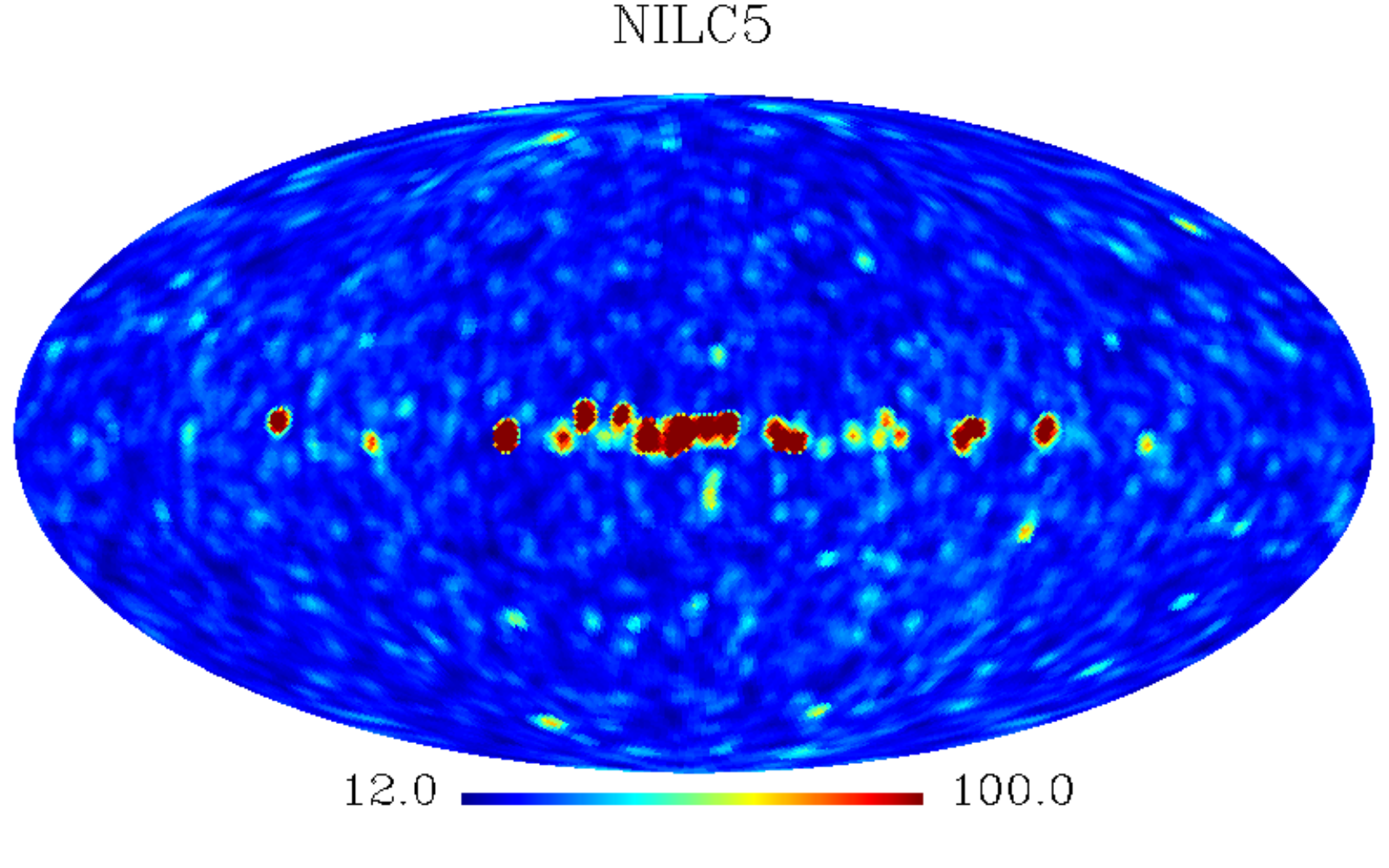}\includegraphics[scale=.15]{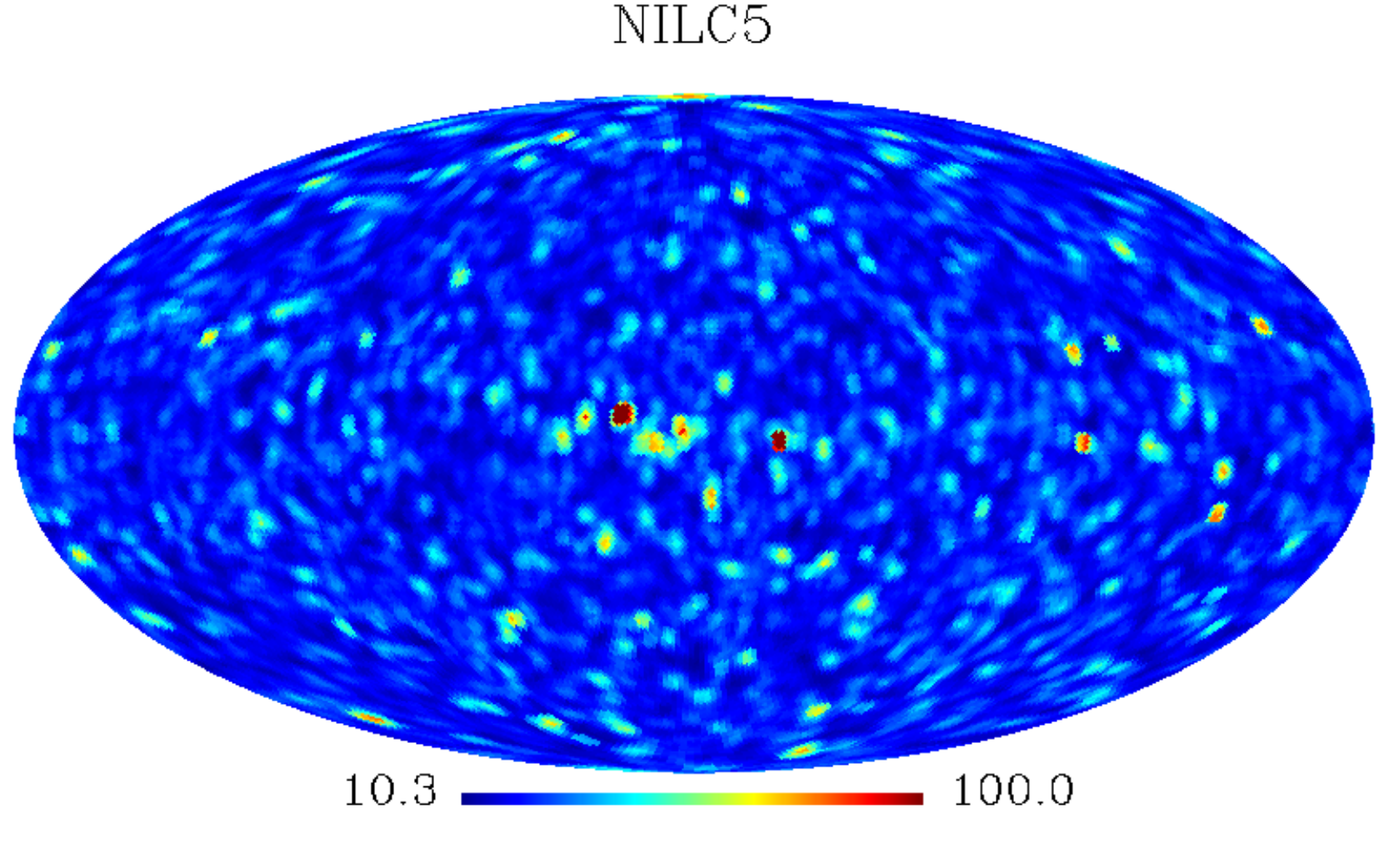}}
    \centerline{\includegraphics[scale=.15]{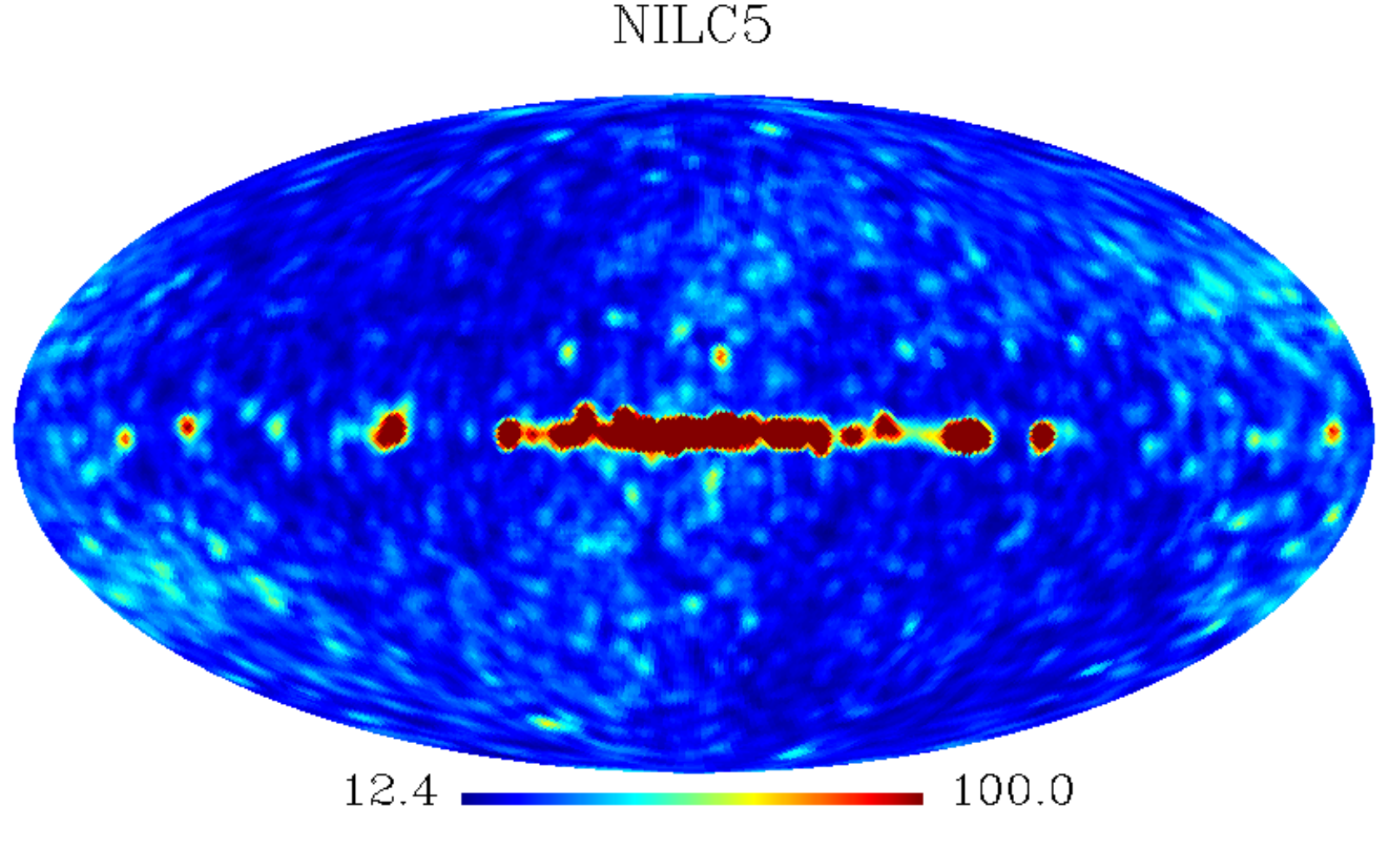}\includegraphics[scale=.15]{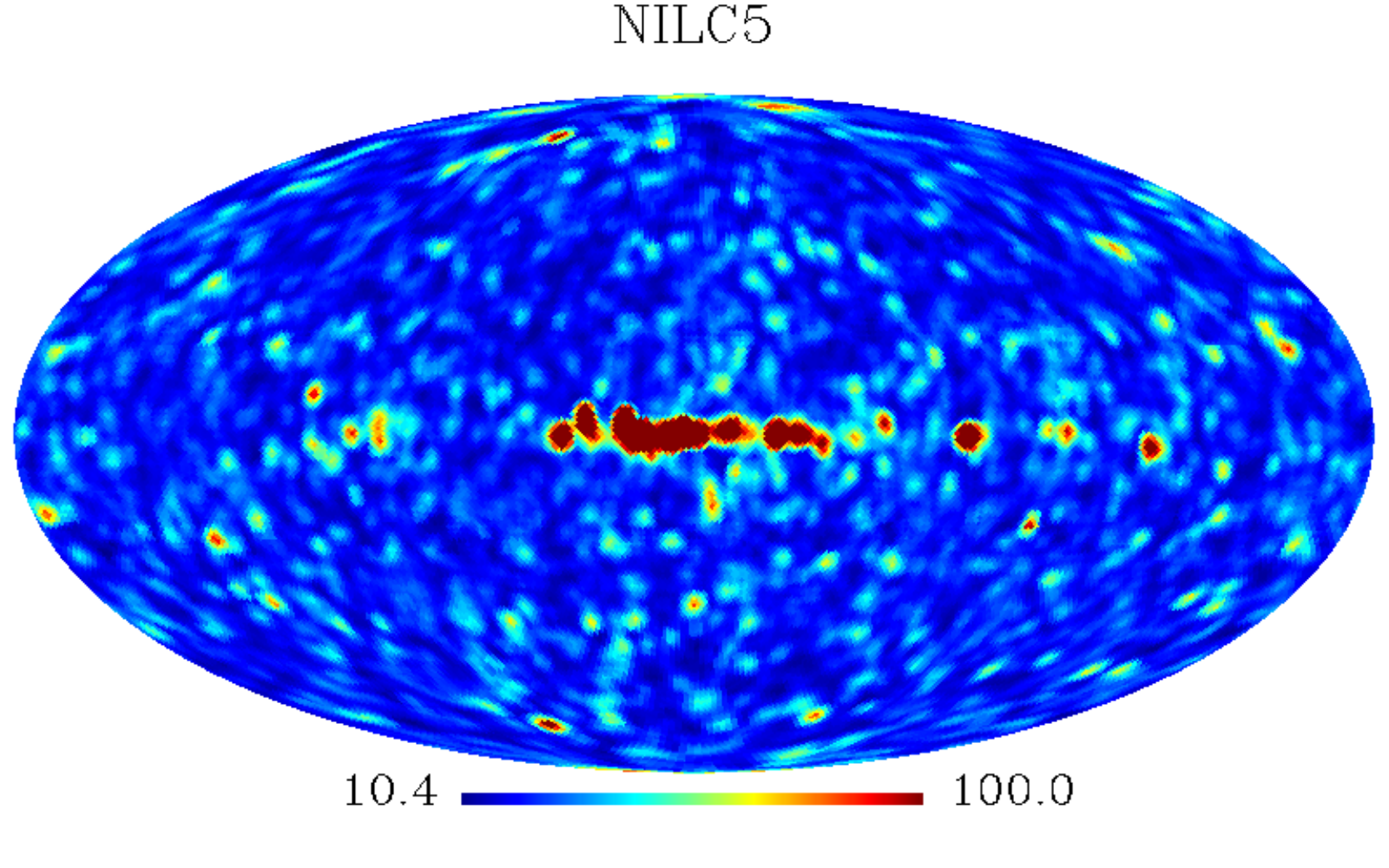}}
    \centerline{\includegraphics[scale=.15]{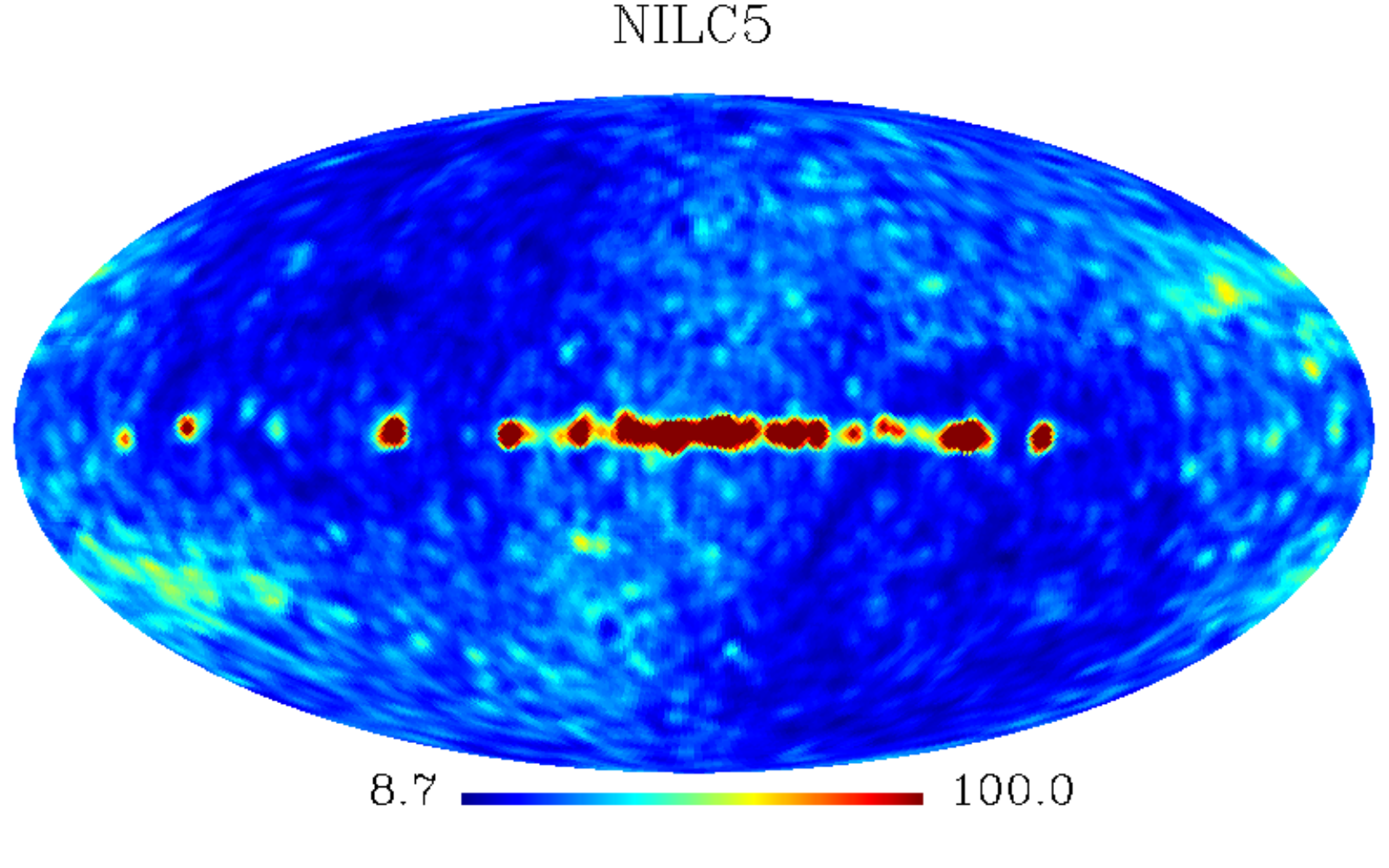}\includegraphics[scale=.15]{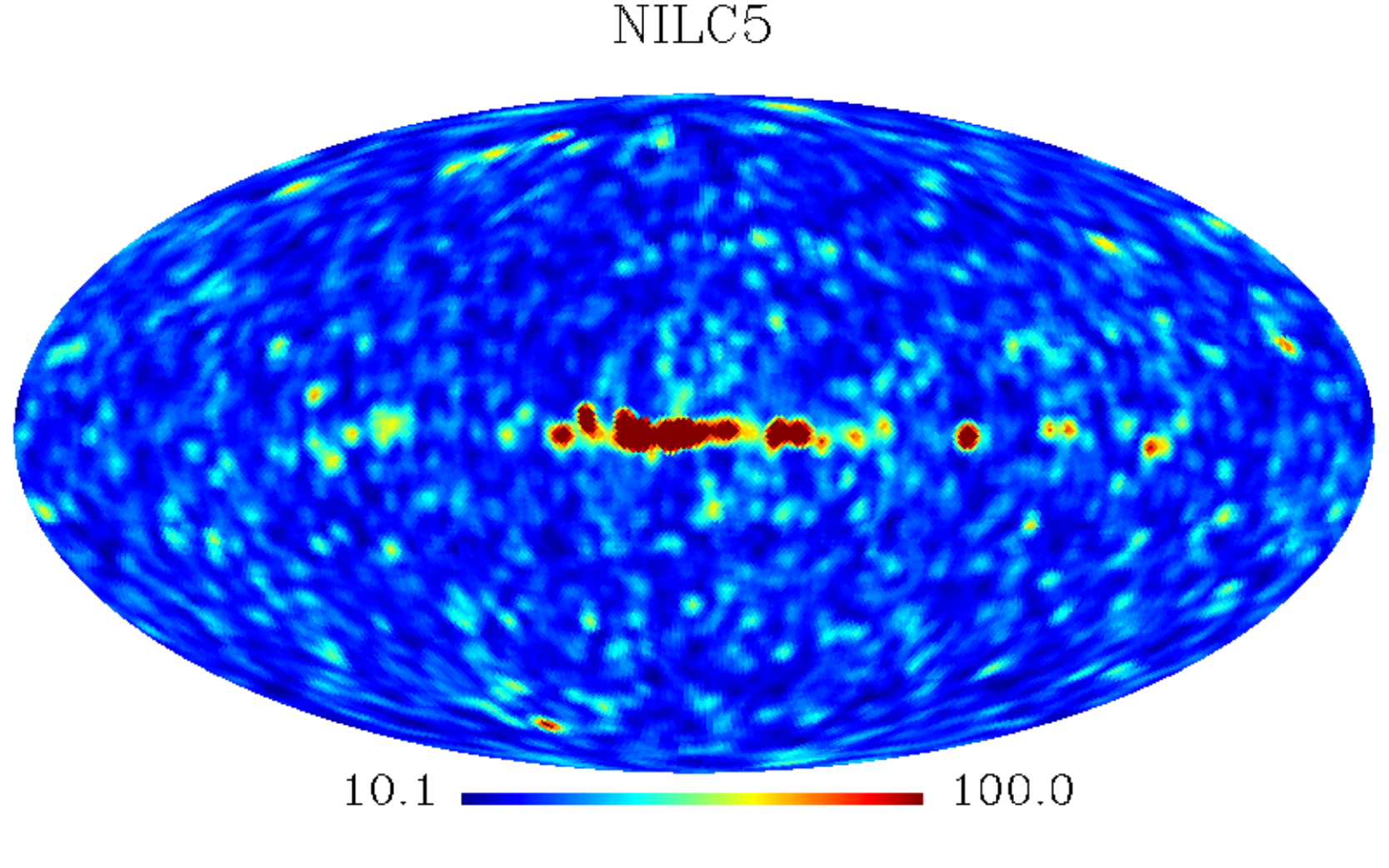}}
    \caption{$\bar{X}_0(R)$ maps (upper), $\bar{X}_1(R)$ maps (middle), and $\bar{X}_2(R)$ maps (lower) for NILC5 data with $R=2^{\circ}$.
    In the left panels, the NILC5 map is smoothed by $\theta_s=10'$, and in right panels, $\theta_s=40'$ smoothing is applied.}
    \label{fig11}
  \end{center}
\end{figure}

Let us turn to NILC5 map, which has the much higher resolution
than ILC7. We firstly study effect of different levels of
smoothing. By adopting $R=2^{\circ}$, in Fig. \ref{fig11} we plot
the $\bar{X}_i$ maps for $\theta_s=10'$ (left panels) and
$\theta_s=40'$ (right panels). Interesting enough for the lower
smoothing case, from the left middle and left lower panels we find
the clear structure of ${\rm N}_{\rm obs}$, i.e. the effective
observations of WMAP for each pixel. We find a smaller ${\rm
N}_{\rm obs}$, which follows a smaller pixel-noise variance,
corresponds a larger $\bar{X}_i^j$. So, these two local MF
statistics can also be used to search for the morphology of the
pixel-noise variance, which has been hidden in the temperature
anisotropy map. We leave this as a future work. Here, in order to
reduce the effect of them, we should choose a larger smoothing
parameter $\theta_s$ as in the right panels in Fig. \ref{fig11},
where we find the morphology of the pixel-noise variance
disappears. But the non-Gaussianity in the Galactic place is still
there, due to the heavy contamination caused by foreground
residuals.

\begin{figure}
  \begin{center}
    \centerline{\includegraphics[scale=.10]{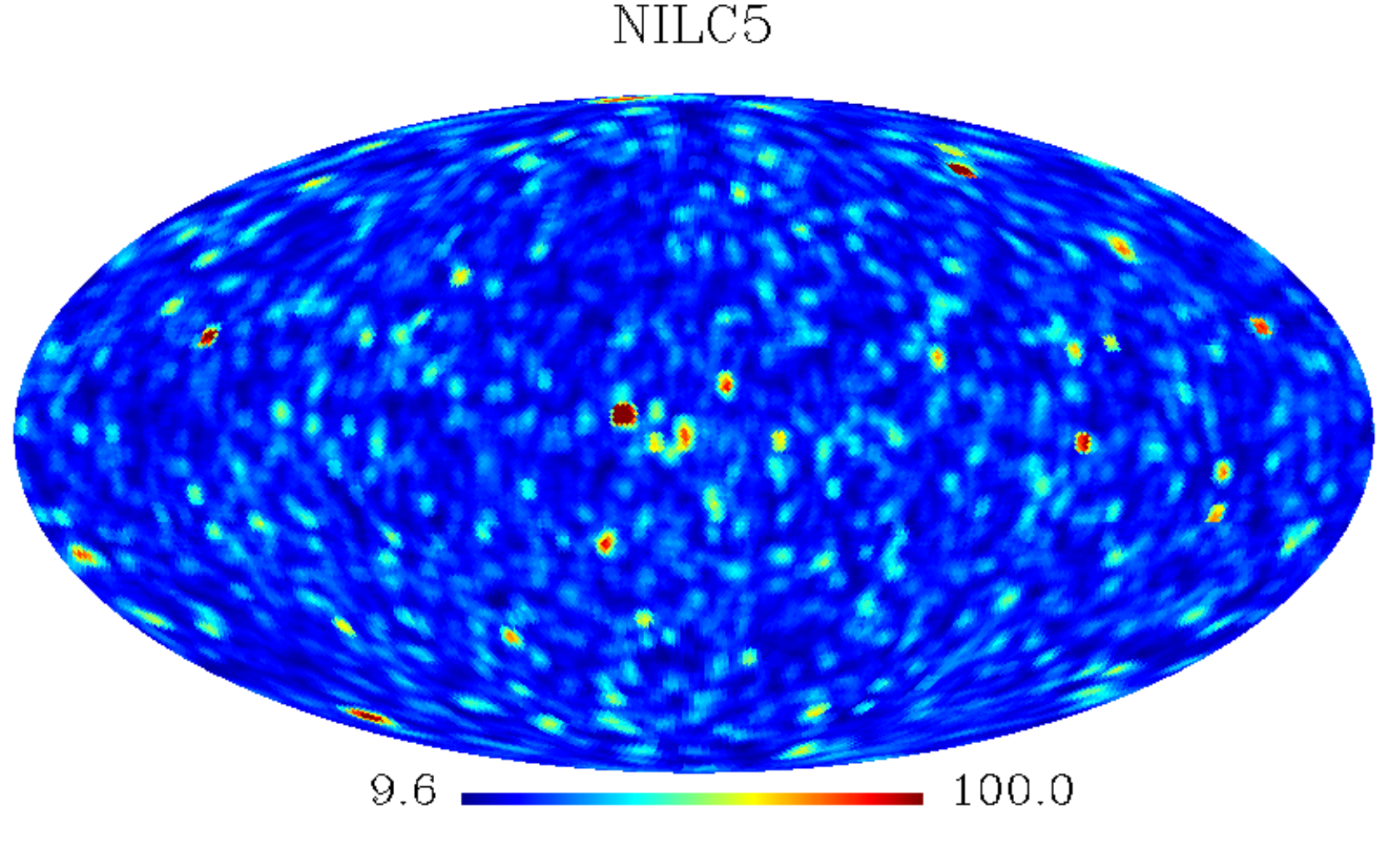}\includegraphics[scale=.10]{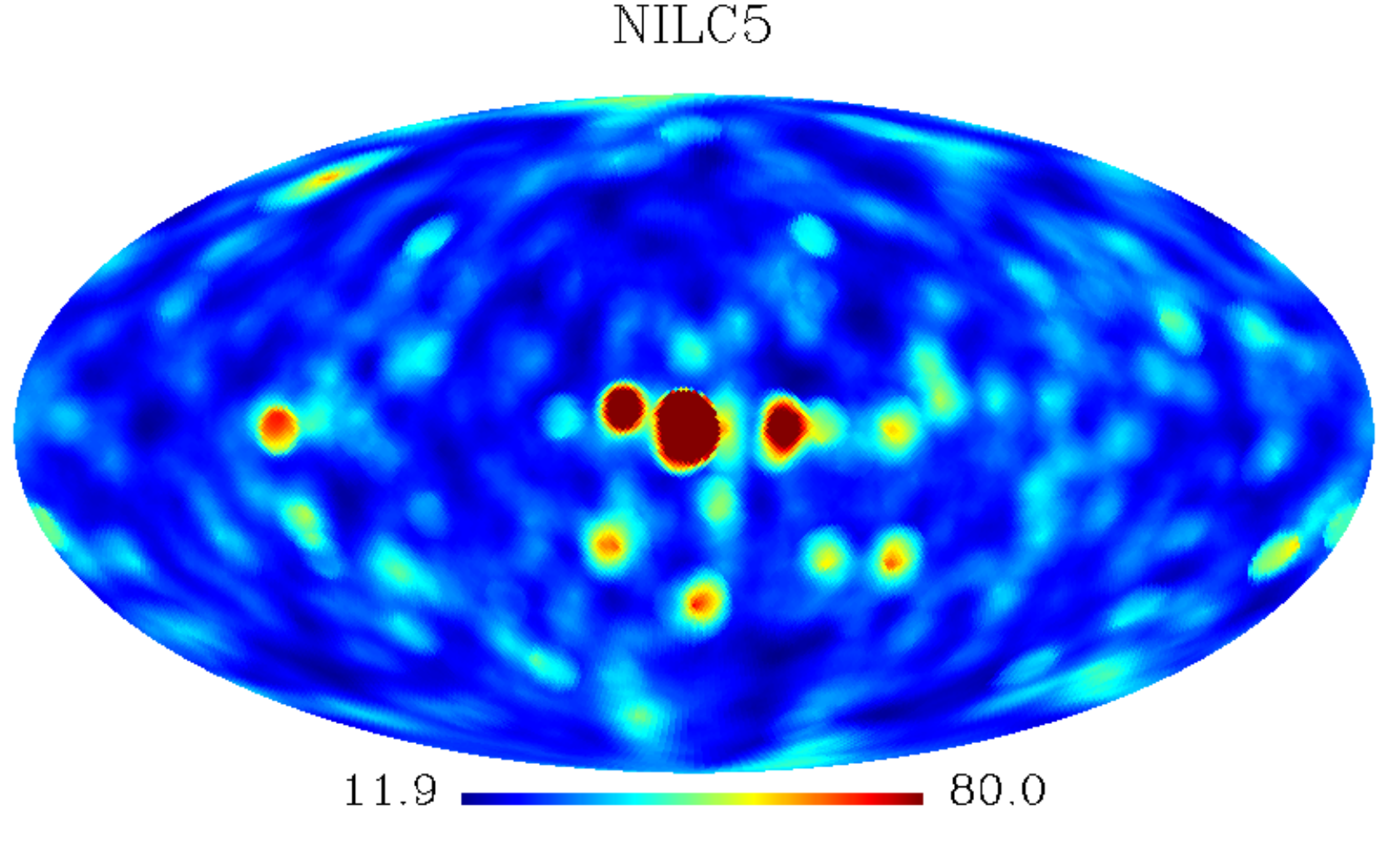}\includegraphics[scale=.10]{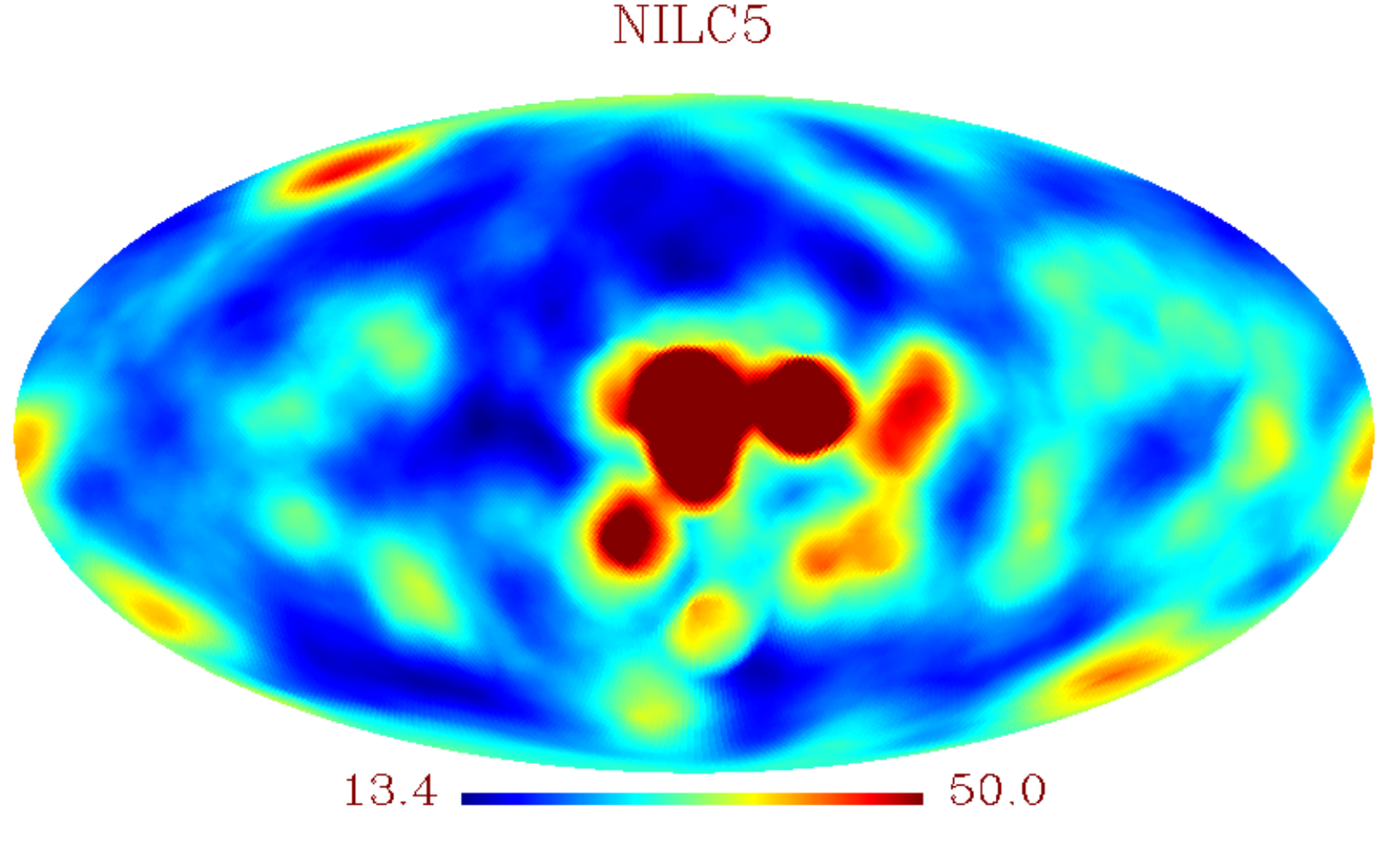}}
    \centerline{\includegraphics[scale=.10]{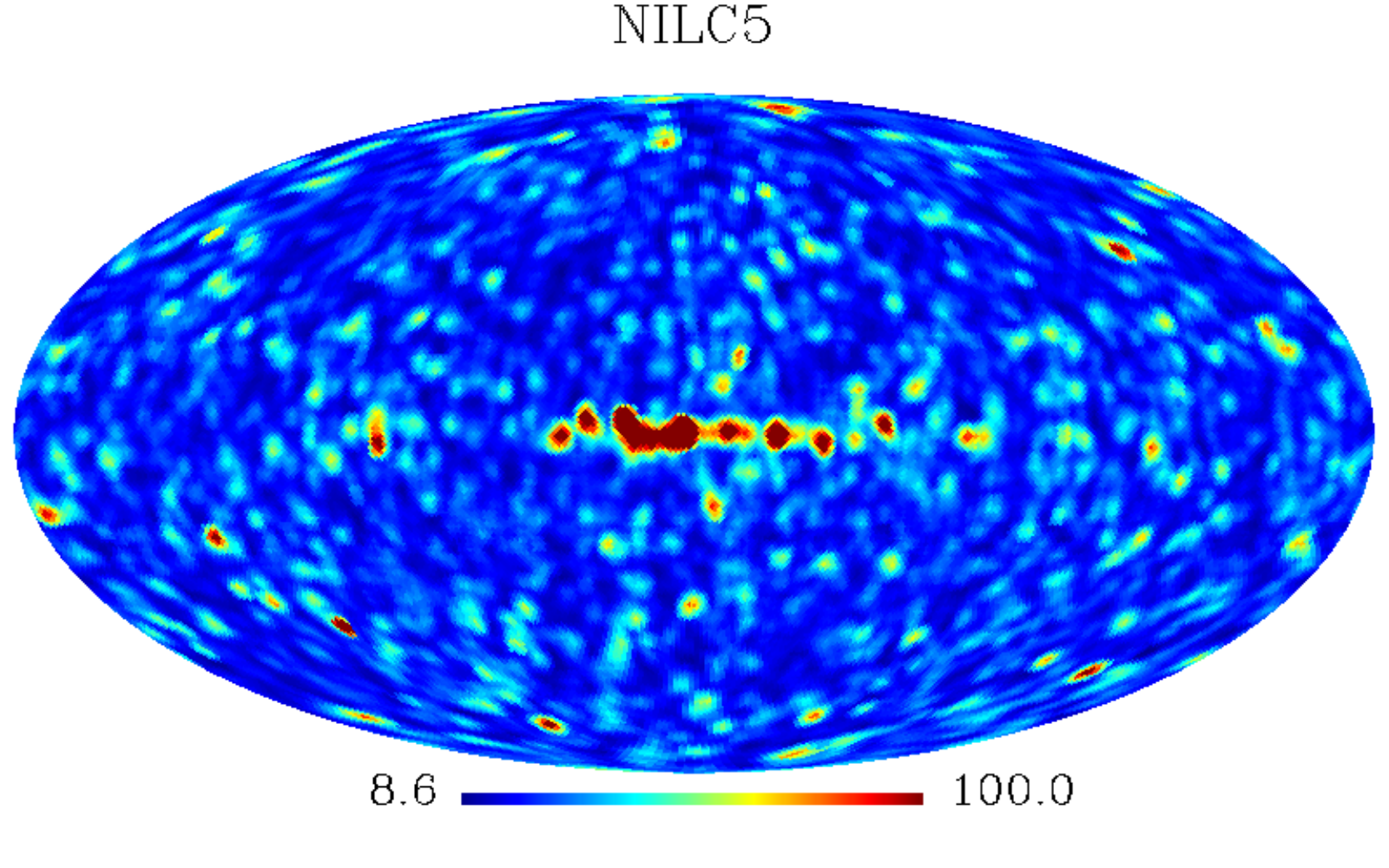}\includegraphics[scale=.10]{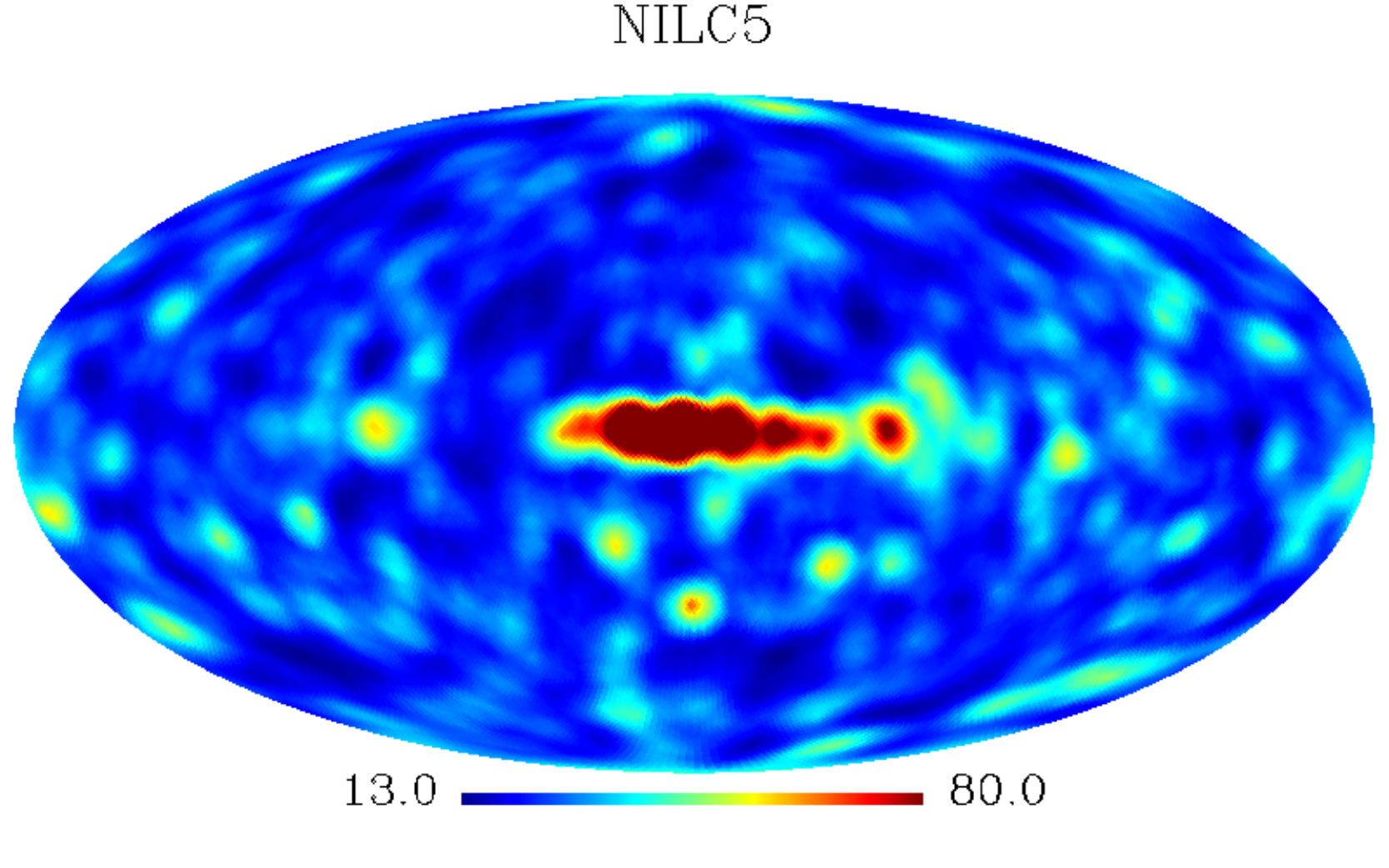}\includegraphics[scale=.10]{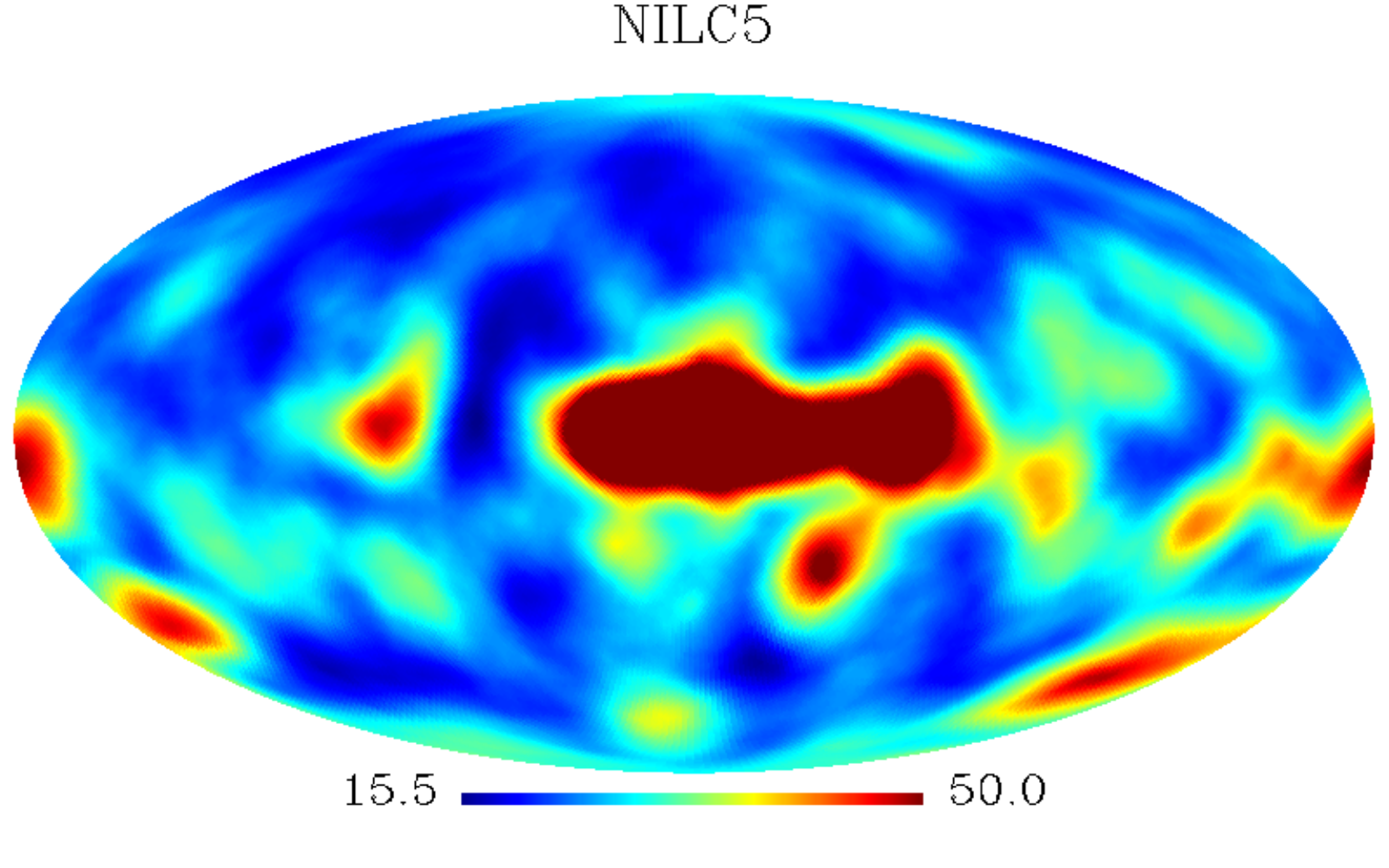}}
    \centerline{\includegraphics[scale=.10]{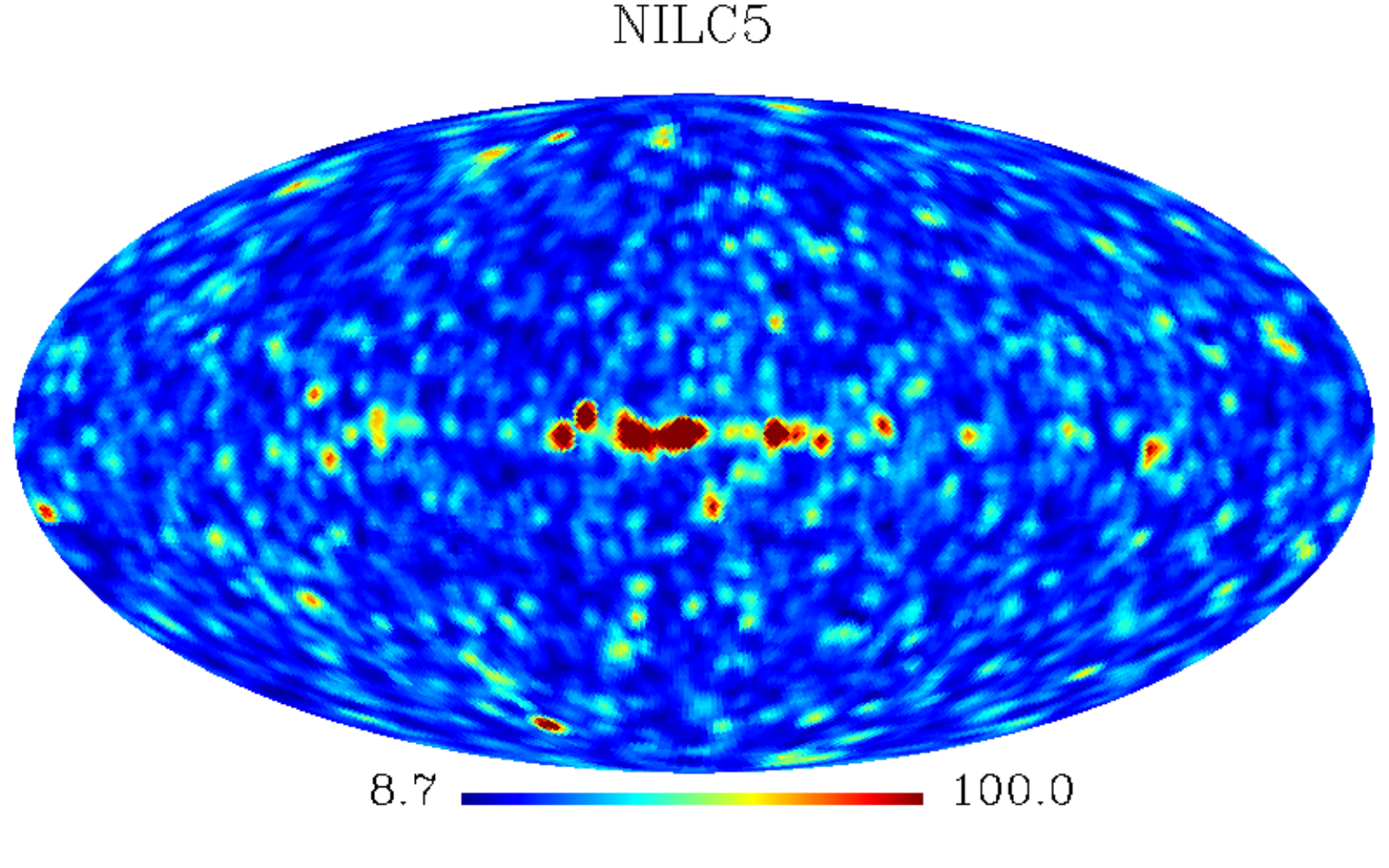}\includegraphics[scale=.10]{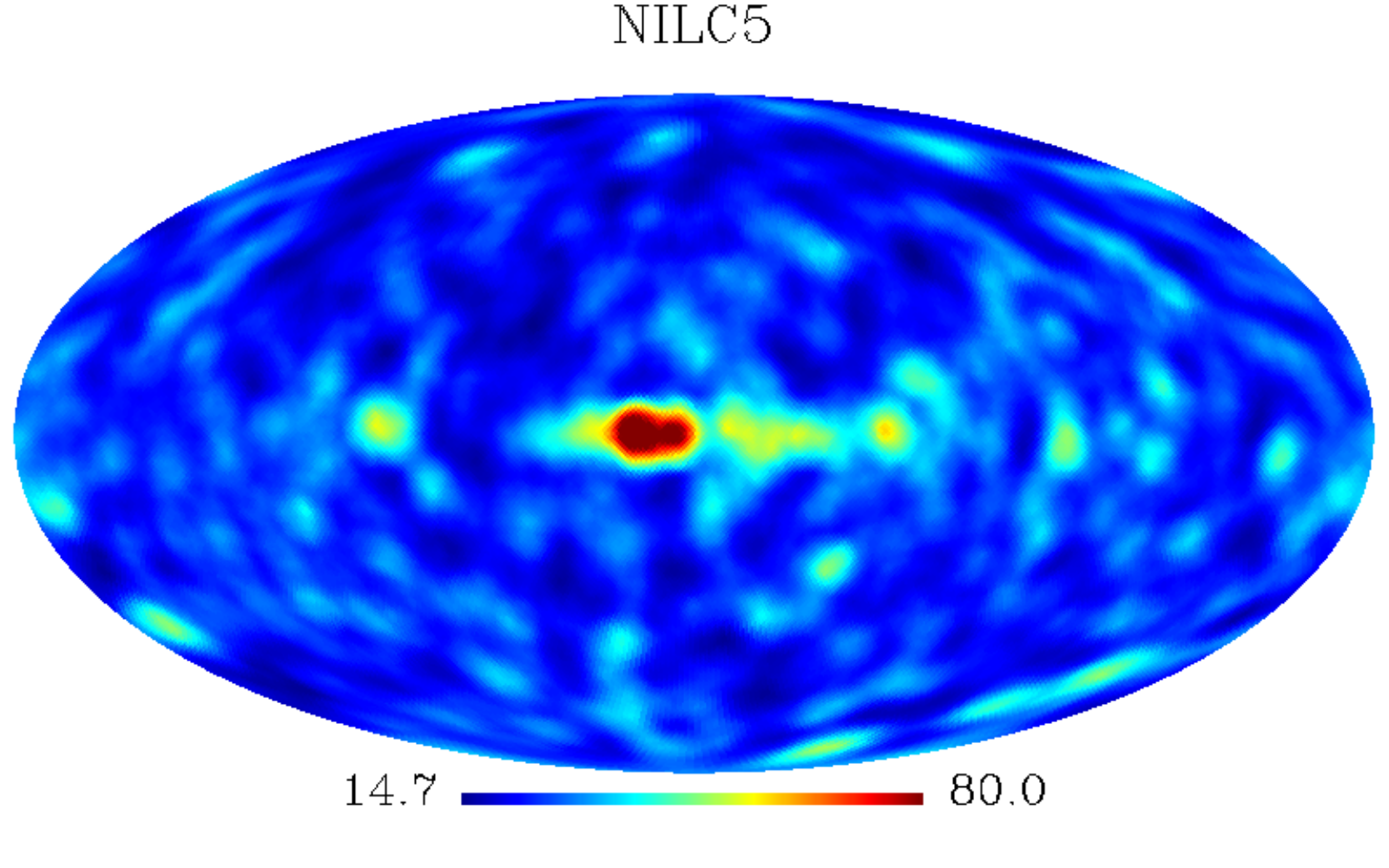}\includegraphics[scale=.10]{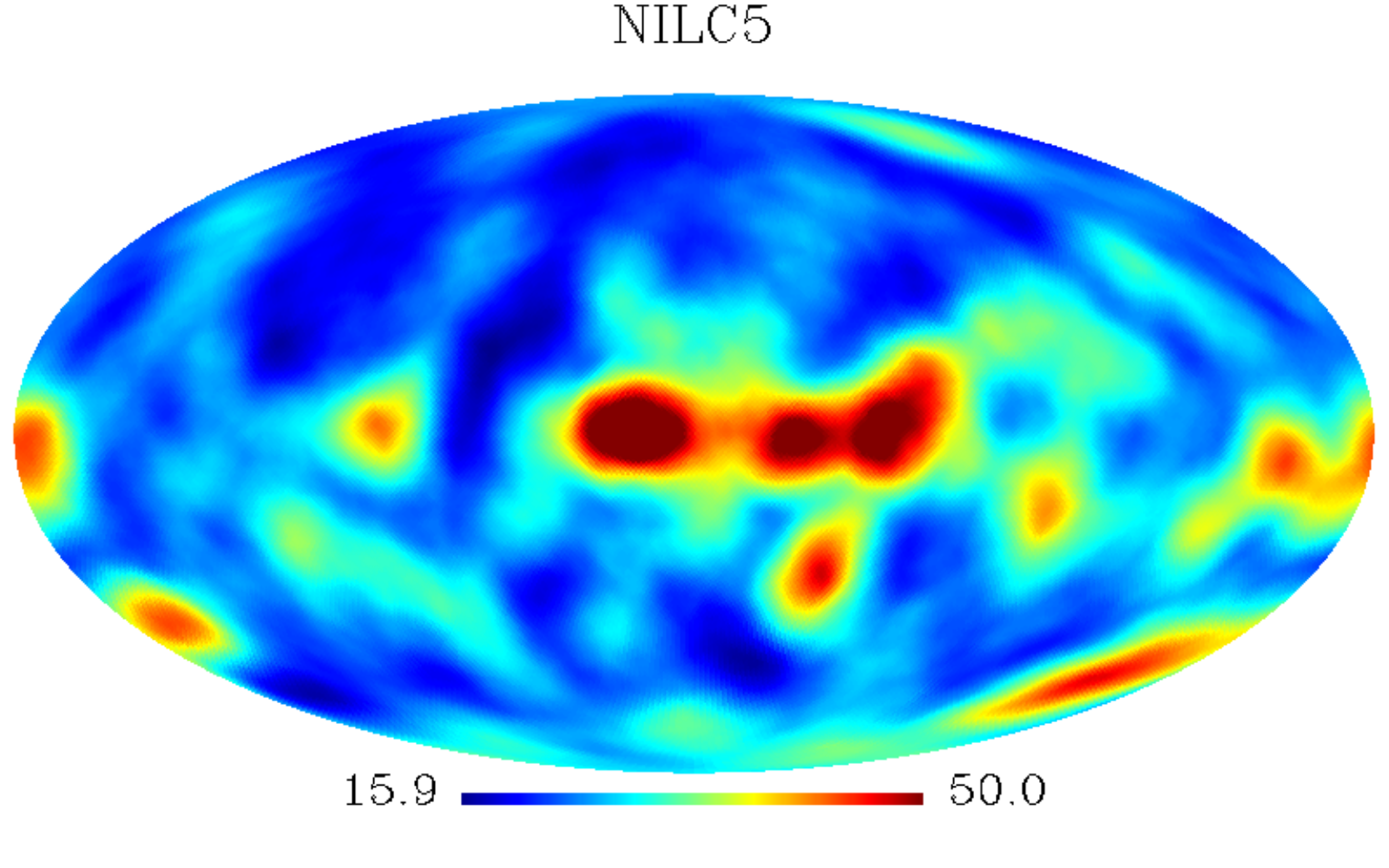}}
    \caption{$\bar{X}_0(R)$ maps (upper), $\bar{X}_1(R)$ maps (middle), and $\bar{X}_2(R)$ maps (lower) for NILC5 data with $\theta_s=60'$ smoothing.
    In the left panels, we have used $R=2^{\circ}$, in middle panels, $R=5^{\circ}$ is chosen, and in right panel, $R=10^{\circ}$.}
    \label{fig111}
  \end{center}
\end{figure}

Choosing a common smoothing parameter $\theta_s=60'$, in Fig.
\ref{fig111} we plot the $\bar{X}_i$ maps for $R=2^{\circ}$
(left), $R=5^{\circ}$ (middle) and $R=10^{\circ}$ (right). By
comparing with those in Fig. \ref{fig1}, we find that although
several non-Gaussian point sources and the foreground residuals in
Galactic plane are still there, NILC5 is much cleaner than ILC7,
as claimed in \cite{delabrouille2009}. Meanwhile, we find that the
non-Gaussianity around WMAP CS are quite significant in the panels
with $R=10^{\circ}$, which will be quantified in next subsection.

\subsection{Local Properties of WMAP Cold Spot}

Now, let us focus on the local properties of WMAP CS, by comparing
with the Gaussian random simulaions. Throughout this section, we
only consider the maps which have been smoothed by $\theta_s=60'$
to exclude the small-scale contaminations.

\begin{figure}
  \begin{center}
    \centerline{\includegraphics[scale=.50]{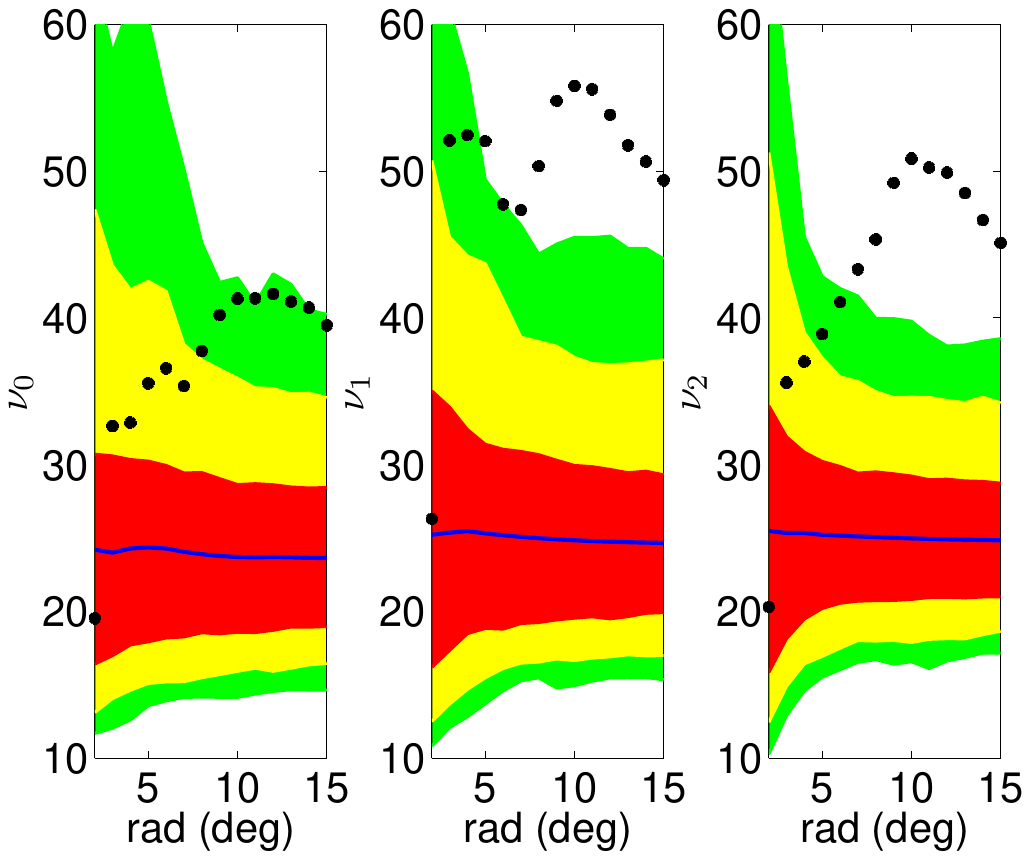}}
    \caption{Three statistics for the spots at ($l=209^{\circ}$, $b=-57^{\circ}$). Confidence regions obtained from 1000 Monte Carlo simulations are shown for 68 pre cent (dark central region, red online), 95 pre cent (lighter middle region, yellow online) and 99 pre cent (lightest outer region, green online) levels, as it the mean (solid blue line). The observed statistics for WMAP ILC7 map are shown by the solid dots (black online).}
    \label{figa}
  \end{center}
\end{figure}

Firstly, we compare WMAP CS with the spots at the same position
($l=209^{\circ}$, $b=-57^{\circ}$) in the random simulations. For
a given $\bar{X}_i(R)$ map ($i=0,1,2$) derived from WMAP data, the
values of $\bar{X}_i(R)$ centered at CS are calculated for the
scales of $R=2^{\circ}, 3^{\circ}, 4^{\circ}, 5^{\circ},
6^{\circ}, 7^{\circ}, 8^{\circ}, 9^{\circ}, 10^{\circ},
11^{\circ}, 12^{\circ}, 13^{\circ}, 14^{\circ}, 15^{\circ}$. The
statistics for ILC7 maps are displayed in Fig. \ref{figa}. We
compare them with 1000 Gaussian simulations. For each simulated
sample, we select the spot at ($l=209^{\circ}$, $b=-57^{\circ}$)
and derive the corresponding $\bar{X}_i(R)$ maps. Then for each
$i$ and $R$, we study the distribution of 1000 $\bar{X}_i(R)$
values, and construct the confident intervals for the statistics.
The $68\%$, $95\%$ and $99\%$ confident intervals are illustrated
in Fig. \ref{figa}. From this figure, we find that for the
$\bar{X}_0(R)$ statistics, WMAP CS is consistent with simulations.
However, for the $\bar{X}_1(R)$ and $\bar{X}_2(R)$ statistics with
$R>6^{\circ}$, WMAP CS deviates from simulation at more than
$99\%$ confident level. For the NILC5 case, we have also obtained
the similar results. These show that as anticipated, WMAP CS is
not a \emph{normal spot}. The deviations at large scales imply
that WMAP CS seems a nontrivial large-scale structure, rather than
a combination of some small non-Gaussian structures (for instance,
the point sources or foreground residuals, which always follow the
non-Gaussianities in the small scales), which is consistent with
the conclusion in \cite{zhao2012}.

\begin{figure}
  \begin{center}
    \centerline{\includegraphics[scale=.50]{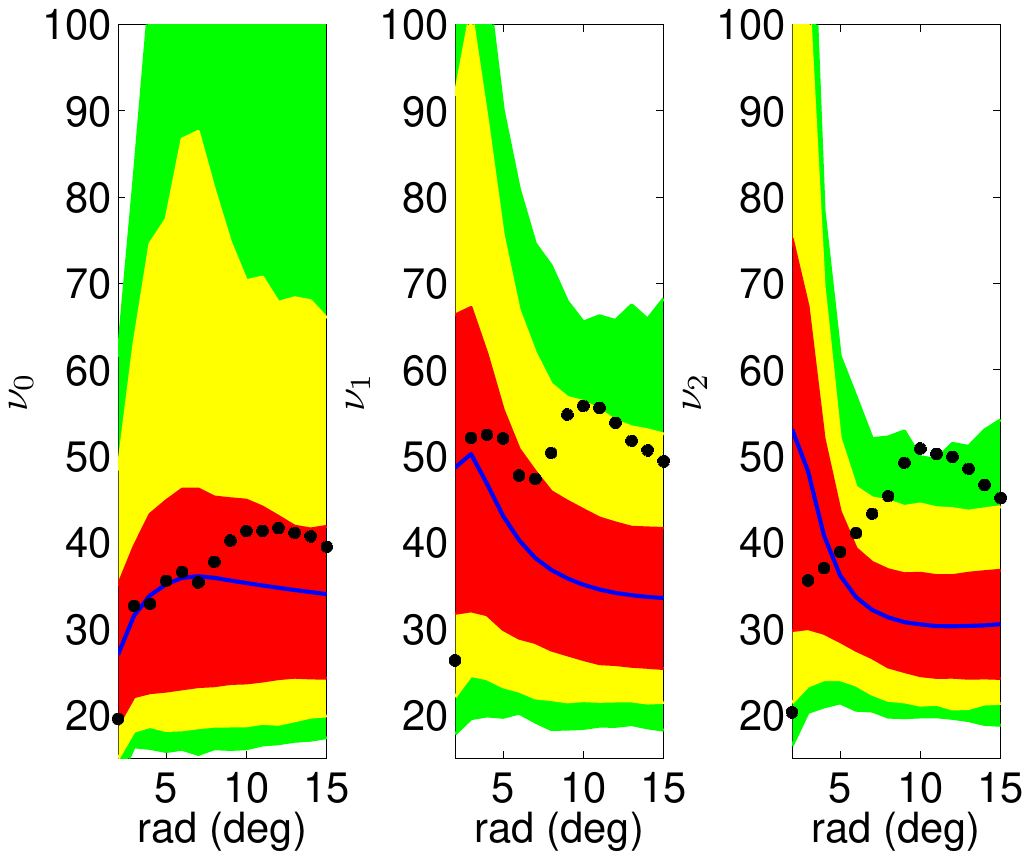}}
    \caption{Three statistics for the \emph{coldest spots}. Confidence regions obtained from 1000 Monte Carlo simulations are shown for 68 pre cent (dark central region, red online), 95 pre cent (lighter middle region, yellow online) and 99 pre cent (lightest outer region, green online) levels, as it the mean (solid blue line). The observed statistics for WMAP ILC7 map are shown by the solid dots (black online).}
    \label{figb}
  \end{center}
\end{figure}

Secondly, we compare WMAP CS with the \emph{coldest spots} in the
Gaussian simulations. For every simulated map with $N_{\rm
side}=512$, we search for the \emph{coldest spot} and derive the
corresponding $\bar{X}_i(R)$ maps by the same calculations used
for WMAP data.  Then for each $i$ and $R$, we study the
distribution of 1000 $\bar{X}_i(R)$ values, and construct the
confident intervals for the statistics. The $68\%$, $95\%$ and
$99\%$ confident intervals are illustrated in Fig. \ref{figb}. We
find that for the MF $\nu_0$, WMAP CS is a normal \emph{coldest
spot} in all the scales when comparing with simulations. For
$\nu_1$ with $R>7^{\circ}$, WMAP CS deviates from the Gaussianity
only at $68\%$ confident level. However, for the MF $\nu_2$ with
$R>7^{\circ}$, the deviation is quite significant (i.e. more than
$95\%$ level). Especially, for $\nu_2$ at the scale
$R=10^{\circ}\sim11^{\circ}$, the deviations are more than $99\%$
confident level. And also in the NILC5 case, the similar
deviations for these statistics have also been derived. So, we
conclude that compared with the \emph{coldest spots} in
simulations, WMAP CS significantly deviates from Gaussianity for
MF $\nu_2$ at the scale $R\sim 10^{\circ}$. However, in the
smaller scales $R<6^{\circ}$, the deviations do not exist. These
support that WMAP CS seems a large-scale non-Gaussian structure.

\begin{figure}
  \begin{center}
    \centerline{\includegraphics[scale=.50]{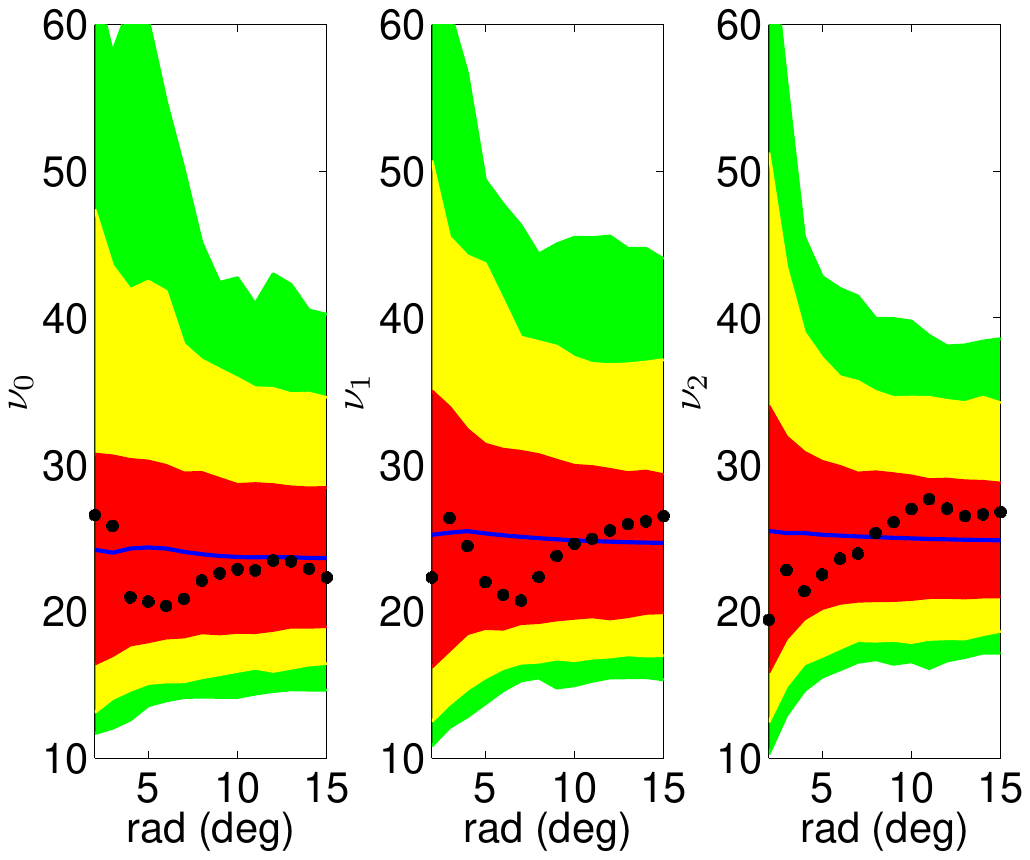}}
    \caption{Same with Fig. \ref{figa}, but here the cosmic texture structure has been subtracted from WMAP ILC7 map.}
    \label{figc}
  \end{center}
\end{figure}

\begin{figure}
  \begin{center}
    \centerline{\includegraphics[scale=.50]{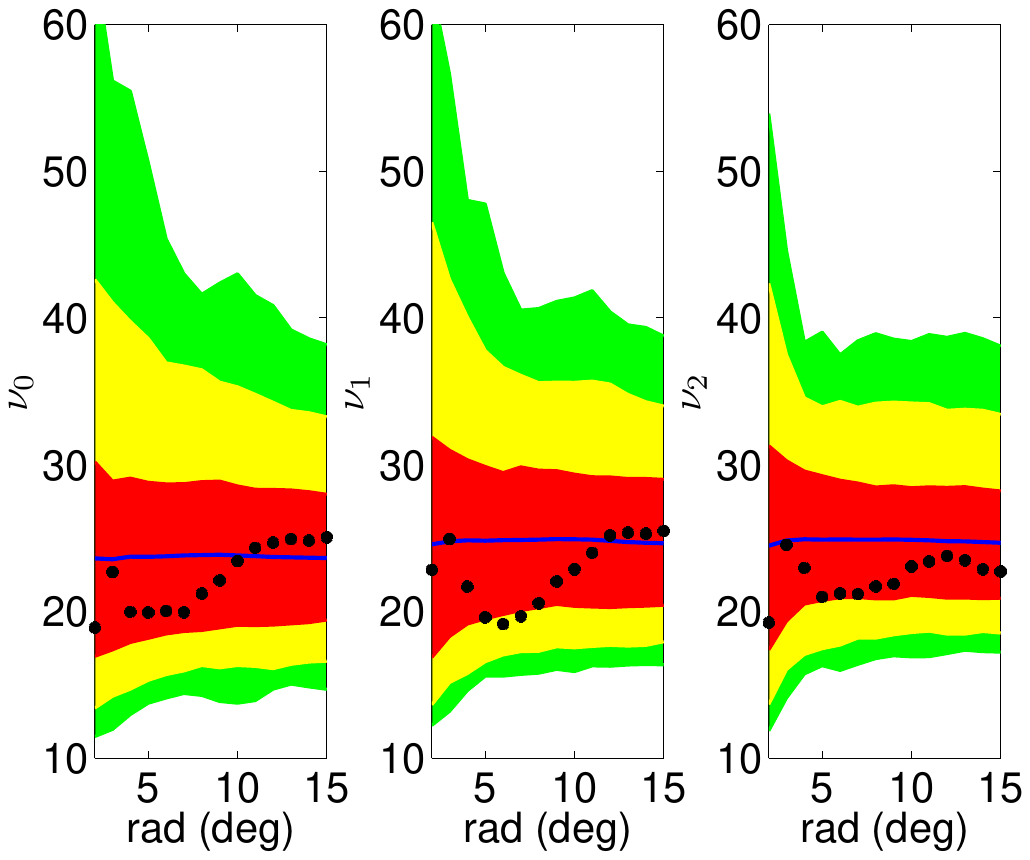}}
    \caption{Same with Fig. \ref{figc}, but here WMAP ILC7 map is replaced by NILC5 map.}
    \label{figd}
  \end{center}
\end{figure}

In \cite{cruz2007b,cruz2008}, the authors found that the cosmic
texture, rather than the other explanations, provides an excellent
interpretation for WMAP CS, which has also been strongly supported
by studying local mean temperature, variance, skewness and
kurtosis in our previous work \cite{zhao2012}. Here we shall test
if the local anomalies of WMAP CS found in MFs are consistent with
the cosmic texture interpretation. The profile for the CMB
temperature fluctuation caused by a collapsing cosmic texture is
given by
\begin{equation}
\label{eq:text_profile2}
\frac{\Delta T}{T} = - \left\{
\begin{array}{ll}
\frac{\varepsilon}{\sqrt{1+4\left(\frac{\vartheta}{\vartheta_c}\right)^2}} & \mathrm{if}~~ \vartheta \leq \vartheta_* \\
&\\
\frac{\varepsilon}{2}\mathrm{e}^{-\frac{1}{2\vartheta^2_c}\left(\vartheta^2 + \vartheta^2_*\right)} & \mathrm{if}~~ \vartheta > \vartheta_* \\
\end{array}
\right.
\end{equation}
where $\vartheta$ is the angle from the center. $\varepsilon$ is
the amplitude parameter, and $\vartheta_c$ is the scale parameter.
$\vartheta_* = \sqrt{3}/2\vartheta_c$. By the Bayesian analysis,
the best-fit texture parameters were obtained
$\varepsilon=7.3\times10^{-5}$ and $\vartheta_c=4.9^{\circ}$
\cite{cruz2007b}. In our calculation, we adopt these best-fit
parameters and subtract this cosmic texture structure from the
ILC7 and NILC5 maps. Then, we repeat the analyses above by using
these subtracted ILC maps. The corresponding results are presented
in Fig. \ref{figc} for ILC7 and Fig. \ref{figd} for NILC5, where
we find that WMAP CS becomes quite normal, i.e. it becomes
excellently consistent with the \emph{normal spots} in Gaussian
simulations for any MF at any scale. So, we obtain the conclusion:
The local analysis of WMAP CS by using the local MF statistics
strongly supports the cosmic texture explanation.

\section{Conclusion}
\label{sec4} Since the release of WMAP data, many attentions have
been paid to study the non-Gaussian CS at Galactic coordinate
($l=209^{\circ}$, $b=-57^{\circ}$), which might be produced by
various small-scale contaminations, such as point sources or
foregrounds, the supervoid in the Universe, or some phase
transition in the early Universe. In this paper, in order to
identify the WMAP CS and discriminate various explanations, we
study the local properties of CS by introducing the local MFs as
the statistics. We find that compared with random Gaussian
simulations, WMAP CS definitely deviates from the \emph{normal
spots} in simulations. In particular, it also deviates from the
\emph{coldest spots} in Gaussian samples at more than $99\%$
confident level at the scale $R\sim 10^{\circ}$. All these support
that WMAP CS is a large-scale non-Gaussian structure. Meanwhile,
we find that the cosmic texture with a characteristic scale about
$10^{\circ}$, which is claimed to the most promising explanation
of CS by many authors, can excellently account for these anomalies
of the local statistics. So our analysis supports the cosmic
texture explanation for WMAP CS.

In the end of this paper, it is important to mention that the
non-Gaussianity of WMAP Cold Spot has been confirmed by the recent
PLANCK observation \cite{planck2013} on the CMB temperature
anisotropies. In the near future, the polarization results of
PLANCK mission will be released, which would play a crucial role
to test WMAP CS and reveal its physical origin \citep{cruz2007b}.

%%%%%%%%%%%%%%%%%%%%%%%%%%%%%%%%%%%%%%%%%%%%%%%%%%%%%%%%%%%%%%%%%%%%%%%%%%%%%%%%%%%
%%%%%%%%%%%%%%%%%%%%%%%%%%%%%%%%%%  Acknowledgments   %%%%%%%%%%%%%%%%%%%%%%%%%%%%%%%%%%%%%%%
%%%%%%%%%%%%%%%%%%%%%%%%%%%%%%%%%%%%%%%%%%%%%%%%%%%%%%%%%%%%%%%%%%%%%%%%%%%%%%%%%%%

{\bf Acknowledgements:} We appreciate useful discussions with P.
Naselsky, J. Kim, M. Hansen and A.M. Frejsel. We acknowledge the
use of the Legacy Archive for Microwave Background Data Analysis
(LAMBDA). Our data analysis made the use of HEALPix \cite{healpix}
and GLESP \cite{glesp}. This work is supported by
project 973 No.2012CB821804, NSFC No.11173021, 11322324 and project of KIP of CAS.

%%%%%%%%%%%%%%%%%%%%%%%%%%%%%%%%%%%%%%%%%%%%%%%%%%%%%%%%%%%%%%%%%%%%%%%%%%%%%%%%%%%
%%%%%%%%%%%%%%%%%%%%%%%%%%%%%%%%%%  BIBLIOGRAPHY  %%%%%%%%%%%%%%%%%%%%%%%%%%%%%%%%%%%%%%%%
%%%%%%%%%%%%%%%%%%%%%%%%%%%%%%%%%%%%%%%%%%%%%%%%%%%%%%%%%%%%%%%%%%%%%%%%%%%%%%%%%%%


\begin{thebibliography}{99}
\bibitem{note2}
Note that, for the WMAP V-band or W-band maps, we always have to
apply the complicated masks (such as the KQ75y7 mask) to exclude
various contaninations casued by foregrounds and dusts. Thus,
there is no way to define the quantity
$\langle{v_i(\nu;R)}\rangle$, and the related quantity $\chi^2$
defined in Eq.(\ref{chi2}). This is the reason why we cannot apply
the method to the WMAP V-band or W-band maps.
\end{thebibliography}
\end{document}